\documentclass[showpacs,twocolumn,aps,psfig,epsf]{revtex4}

\usepackage{graphics}% Include figure files
\usepackage{dcolumn}% Align table columns on decimal point
\usepackage{bm}% bold math

\usepackage{epsfig}

\begin{document}

\newcommand{\vAi}{{\cal A}_{i_1\cdots i_n}}
\newcommand{\vAim}{{\cal A}_{i_1\cdots i_{n-1}}}
\newcommand{\vAbi}{\bar{\cal A}^{i_1\cdots i_n}}
\newcommand{\vAbim}{\bar{\cal A}^{i_1\cdots i_{n-1}}}
\newcommand{\htS}{\hat{S}}
\newcommand{\htR}{\hat{R}}
\newcommand{\htB}{\hat{B}}
\newcommand{\htD}{\hat{D}}
\newcommand{\htV}{\hat{V}}
\newcommand{\cT}{{\cal T}}
\newcommand{\cM}{{\cal M}}
\newcommand{\cMs}{{\cal M}^*}
\newcommand{\vk}{{\vec k}}
\newcommand{\vK}{{\vec K}}
\newcommand{\vb}{{\vec b}}
\newcommand{{\vp}}{{\vec p}}
\newcommand{{\vq}}{{\vec q}}
\newcommand{\vQ}{{\vec Q}}
\newcommand{\vx}{{\vec x}}
\newcommand{\tr}{{{\rm Tr}}}
\newcommand{\beq}{\begin{equation}}
\newcommand{\eeq}[1]{\label{#1} \end{equation}}
\newcommand{\half}{{\textstyle \frac{1}{2}}}
\newcommand{\gton}{\stackrel{>}{\sim}}
\newcommand{\lton}{\mathrel{\lower.9ex
                  \hbox{$\stackrel{\displaystyle <}{\sim}$}}}
\newcommand{\ee}{\end{equation}} \newcommand{\ben}{\begin{enumerate}}
\newcommand{\een}{\end{enumerate}} \newcommand{\bit}{\begin{itemize}}
\newcommand{\eit}{\end{itemize}} \newcommand{\bc}{\begin{center}}
\newcommand{\ec}{\end{center}} \newcommand{\bea}{\begin{eqnarray}}
\newcommand{\eea}{\end{eqnarray}}
\newcommand{\beqar}{\begin{eqnarray}}
\newcommand{\eeqar}[1]{\label{#1} \end{eqnarray}}
\newcommand{\bra}[1]{\langle {#1}|}
\newcommand{\ket}[1]{|{#1}\rangle}
\newcommand{\norm}[2]{\langle{#1}|{#2}\rangle}
\newcommand{\brac}[3]{\langle{#1}|{#2}|{#3}\rangle}
\newcommand{\hilb}{{\cal H}}
\newcommand{\pleft}{\stackrel{\leftarrow}{\partial}}
\newcommand{\pright}{\stackrel{\rightarrow}{\partial}}

\title{ Non-Abelian energy loss in cold nuclear matter }

\author{Ivan~Vitev}
\email{ivitev@lanl.gov}

\affiliation{ Los Alamos National Laboratory, 
  Theoretical Division and Physics Division, 
  Los Alamos, NM 87545, USA  }

\begin{abstract}
We use a formal recurrence relation approach to multiple parton scattering 
to find the complete solution to 
the problem of medium-induced gluon emission from partons propagating 
in cold nuclear matter. The differential bremsstrahlung spectrum, where 
Landau-Pomeranchuk-Migdal destructive interference effects are fully 
accounted for, is calculated for three different cases: 
(1) a generalization of the incoherent Bertsch-Gunion solution for 
asymptotic on-shell jets, (2) initial-state energy loss of incoming jets 
that undergo hard scattering and (3) final-state energy loss of jets 
that emerge out of a hard scatter. Our analytic solutions are given as an 
infinite opacity series, which represents a cluster expansion of the 
sequential multiple scattering. These new solutions allow, for the first 
time, direct comparison between initial- and final-state energy loss 
in cold nuclei.  We demonstrate that, contrary to the naive assumption,
energy loss in cold nuclear matter can be large. Numerical results to 
first order in opacity show that, in the limit of large jet energies, 
initial- and final-state energy losses exhibit different path 
length dependences, 
linear versus quadratic, in contrast to earlier findings. In addition, 
in this asymptotic limit, initial-state energy loss is considerably 
larger than final-state energy loss. These new results have significant 
implications for heavy ion phenomenology in both p+A and A+A reactions. 
\end{abstract}


\maketitle

\section{Introduction}

Non-Abelian final-state medium-induced radiative energy loss
in the quark-gluon plasma (QGP) is the best studied 
many-body perturbative QCD (pQDC) application for high energy
nuclear collisions. Several theoretical approaches 
that address this question have been well documented in recent
reviews~\cite{Gyulassy:2003mc,Kovner:2003zj,Baier:2000mf,Vitev:2004bh,Majum:2006}.
In contrast, initial-state energy loss in cold nuclear matter,
pertinent to hard jet and particle production in proton-nucleus 
(p+A) and nucleus-nucleus (A+A) collisions, has not been studied
so far.

Strong motivation for in-depth investigation of cold nuclear
matter energy loss comes from the finding that at most 
$1/2$ of the forward rapidity suppression in d+Au collisions
measured at the Relativistic Heavy Ion Collider 
(RHIC)~\cite{Arsene:2004ux,Adams:2006uz}  
can be accounted for by leading twist~\cite{Guzey:2004zp} 
or high twist~\cite{Qiu:2004da} shadowing  calculations 
that take into account constraints from deep inelastic 
scattering (DIS) on nuclei. 
Moreover, at much lower center of mass energies at the 
Super Proton Synchrotron (SPS), a similar forward rapidity 
suppression is established and shown to not be compatible 
with shadowing calculations~\cite{Vitev:2006bi}.

Energy loss for asymptotic $t = \pm \infty $ on-shell jets
that {\em do not} undergo hard scattering was discussed 
in~\cite{Gunion:1981qs} for nuclear matter of extent 
$L \sim  r_0 = 1.2$~fm.  It has been argued that the inclusion 
of the Landau-Pomeranchuk-Migdal (LPM)
destructive interference effect~\cite{Landau:1953,Migdal:1956tc} 
for this  regime, 
in the approximation of infinitely large number of soft scatterings,  
leads to negligible $\Delta E_{\rm rad}$, $1/3$ of the magnitude 
of final-state energy loss~\cite{Baier:2000mf}. Final-state
energy loss in cold nuclei ~\cite{Gyulassy:2003mc} was found to
have the same qualitative behavior as final-state energy 
loss in the QGP, though much smaller in magnitude and with  
possible relevance to suppressed hadron production in 
{\em semi-inclusive} DIS. None of these regimes can yield 
significant and phenomenologically relevant contribution to 
cold nuclear matter attenuation at collider energies.

Thus, the primary purpose of this manuscript is to derive and 
properly compare the Bertsch-Gunion, initial- and final-state energy 
loss in cold nuclei. The dominant contribution to  $\Delta E_{\rm rad}$ 
can  then  be used in heavy ion QCD phenomenology~\cite{prep}. The 
secondary goal of this paper is to clarify the 
principle difference between 
radiative and collisional~\cite{Mustafa:2004dr,Adil:2006ei} energy loss in 
cold/hot nuclear matter. Recent studies have made comparisons
between $\Delta E_{\rm rad}$ and  $\Delta E_{\rm col}$,
emphasizing a deep LPM cancellation regime, which is not representative
of the process of radiative energy loss. In addition, different 
and often incompatible formalisms are used in such comparisons.

We start by recalling the energy loss results for electrodynamics 
(QED)~\cite{jackson}.  Consider medium of atomic density $\rho = N/V$. 
Each atom has  $Z$ electrons of  electric charge $e$ and mass $m$. An 
incident particle with electric charge $ze$ and mass $M$, $E = \gamma M$, 
$p = \beta \gamma M$, undergoes multiple Coulomb scattering in 
such a  medium. Its collisional energy loss per unit length, 
including  momentum transfers down to the mean excitation 
energy of the 
electrons $\langle \omega \rangle$, is given by: 
\beq
\frac{d \Delta E_{coll}}{dx} \approx  64 
\pi^3  \alpha_{em}^2  z^2 Z \rho \frac{1}{\beta^2 m} \ln B_q \;.
\eeq{colQED}
In Eq.~(\ref{colQED})  $B_q = 2 \gamma^2 \beta^2 m / 
\langle \omega \rangle $ and in the high energy limit, 
$\ln B_q \gg \beta^2$, we have neglected a small correction 
to the large logarithm 
related to the relativistic electron spin. 
The physics behind collisional energy loss is the energy,
$\Delta E_{\rm col}(Q^2)$, transfered {\em to} 
the medium by the incident fast  particle. Thus, 
one expects little dependence on the mass of the 
the incident particle, $M$, but strong dependence on the mass of 
the target scatterers, $m$. It is evident from Eq.~(\ref{colQED}) 
that collisional energy loss grows at most logarithmically 
with the energy of the incident particle, $E$, and linearly 
with  the size of the medium, $L$.

On the other hand, the radiative energy loss per unit 
length is given by~\cite{jackson}: 
\beq
\frac{d \Delta E_{rad}}{dx} \approx \frac{1024}{3}
 \pi^3 \alpha_{em}^3  
z^4 Z^2 \rho \frac{1}{M^2} E  \ln (\gamma \lambda) \;,
\eeq{radQED}  
where $ \lambda = \lambda(M,\gamma, \omega_s) $ may also
depend on the screening of the target electric charge 
via the characteristic frequency $\omega_s$.
Radiative energy loss arises from the acceleration of the 
incident particle, which allows for emission of real 
photons. Thus, one expects that there will be significant
dependence on the particle mass, $M$. Emitted photons naturally
carry a {\em fraction} of the incident  particle energy, 
implying that radiative energy loss  is proportional to 
$E$ or, equivalently, $\gamma$. Both features are easily 
seen in Eq.~(\ref{radQED}) and $\Delta E_{\rm rad}$ grows 
linearly with $L$.

By comparing Eqs.~(\ref{colQED}) and (\ref{radQED}) one
observes that in the high energy regime $ \Delta E_{\rm rad}
\gg \Delta E_{\rm col} $. Thus, for 
ultra-relativistic particles radiative energy loss is the 
dominant mechanism of momentum attrition. As we will show
below,  the same energy dependence of  $\Delta E_{\rm rad}$ 
is characteristic of QCD energy loss. It is only in the deep 
LPM regime for final-state energy loss that the energy 
dependence of $\Delta E_{\rm rad}$ is reduced to logarithmic.
Since this is a very specific case, care should be
taken to evaluate collisional energy loss in the same
model of momentum transfers with the medium to the same
power in the expansion in $1/p^+$.

This paper is organized as follows: in Section~II we 
review the recursive approach to multiple parton 
scattering in both cold and hot nuclear matter, formulated 
in~\cite{Gyulassy:2000er,Gyulassy:2000fs}. We use final-state 
energy loss as an example. Next, we derive to all orders in opacity  
the two new solutions for the radiative non-Abelian energy loss 
of incoming  partons that may or may not  undergo hard 
scattering that produces high-$p_T$ or high-$E_T$ particles or jets. 
Section III contains a detailed numerical study to first order in 
opacity  of  the three different energy loss regimes. We identify 
initial-state energy loss as the dominant cold nuclear matter 
$\Delta E_{\rm rad}$ contribution, relevant to p+A and 
A+A collider phenomenology.
A summary and discussion of our results is given in Section IV.   
Appendixes contain useful kinematic simplifications, which allow
for the derivation of the  all-orders in opacity solutions given 
here. Explicit brute force calculations to second order in opacity  
of the new Bertsch-Gunion and initial-state $\Delta E_{\rm rad}$ 
are also shown for the purpose of validating the general 
results.

%%%%%%%%%%%%%%%%%%%%%%%%%%%%%%%%%%%%%%%%%%%%%%%%%%%%%%%%%%%%%%%

\begin{figure}[!tb]
\begin{center}
\epsfig{file = 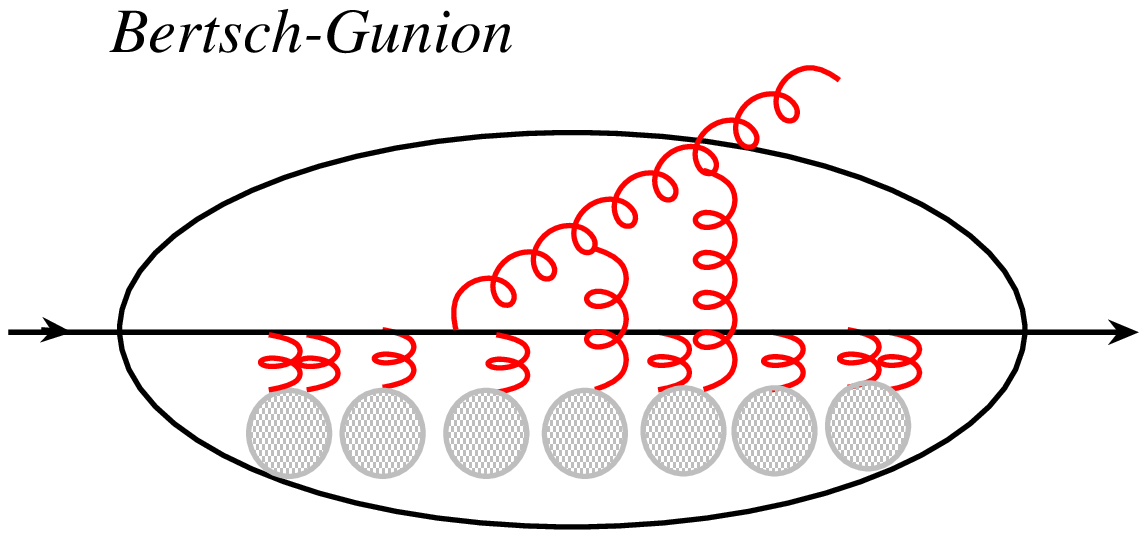,width=1.6in,angle=0}
\epsfig{file = 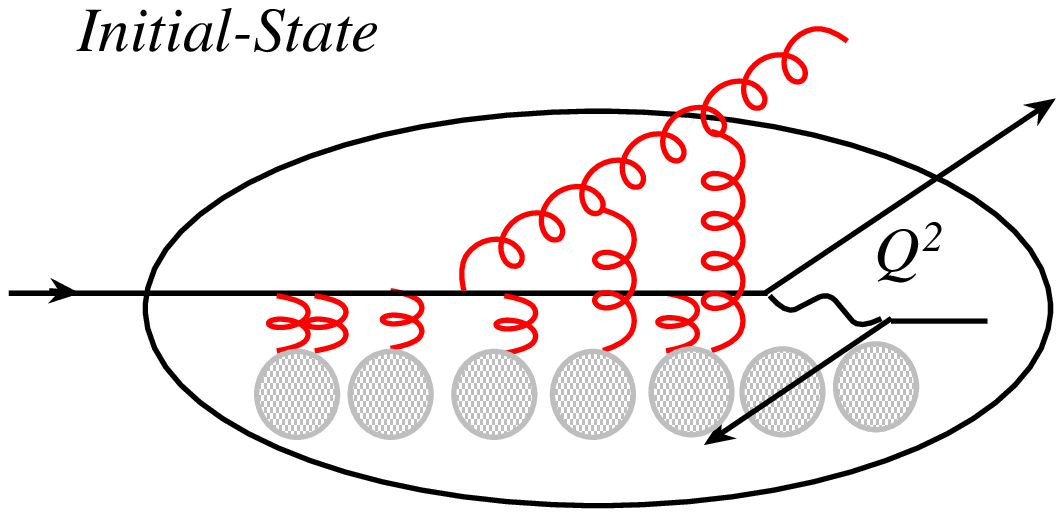,width=1.6in,angle=0}
\epsfig{file = 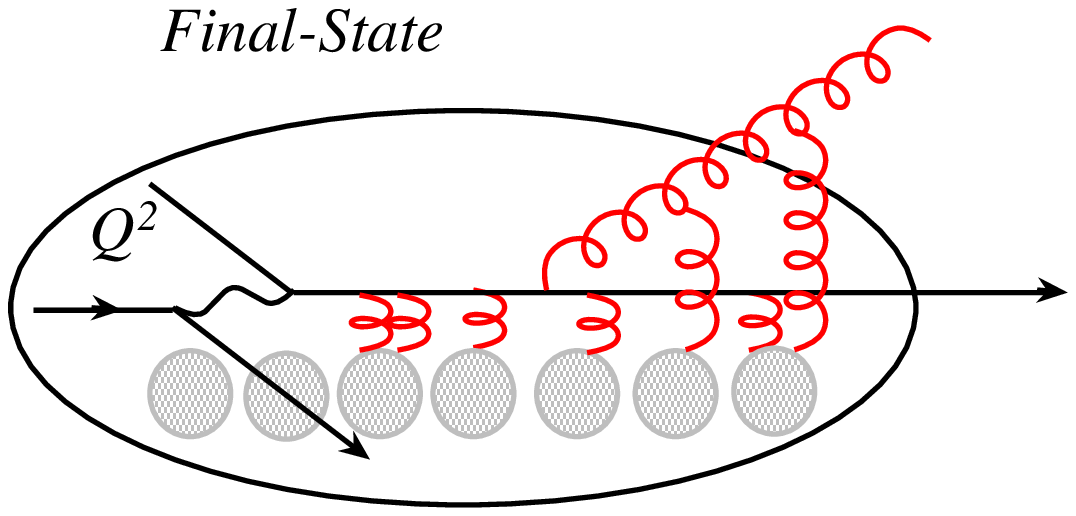,width=1.6in,angle=0}
\vspace*{.1in}
\caption{(Color online) Three distinct cases of medium-induced bremsstrahlung 
are illustrated: 1) Bertsch-Gunion case of $ t=-\infty,    
t = + \infty $   on-shell jets; 2) initial-state  energy loss
in the nucleus followed by a large $Q^2 $ process,
resulting in the production of high-$p_T$ or high-$E_T$ 
particles and jets; 3)  final-state  energy loss in the
nucleus, after a hard scatter. }
\label{fig-schem}
\end{center} 
\end{figure}

%%%%%%%%%%%%%%%%%%%%%%%%%%%%%%%%%%%%%%%%%%%%%%%%%%%%%%%%%%%%%%%%%

\section{Recursive method for  opacity expansion}

For radiative processes in QCD, including medium-induced bremsstrahlung, 
it is important to keep track of the evolution of the gluon transverse
momentum ${\bf k}$ in a plane perpendicular to the direction of 
jet propagation~\cite{Gunion:1981qs}. Such a ${\bf k}$ 
may arise from single hard or multiple soft scattering. The acceleration 
of the color charges in the 2D transverse plane generates 
color currents whose detailed interference pattern determines 
the  strength of the non-Abelian 
Landau-Pomeranchuk-Migdal~\cite{Landau:1953,Migdal:1956tc}
effect. Let us denote by: 
\beqar
\label{hprop}
{\bf H}&=&{{\bf k} \over {\bf k}^2 }\; , \qquad \qquad  \\[1.ex]
{\bf C}_{(i_1 \cdots i_m)}&=&{{\bf k} - {\bf q}_{i_1} - 
\cdots -{\bf q}_{i_m}  
\over ({\bf k} - {\bf q}_{i_1} - 
\cdots -{\bf q}_{i_m}   )^2 } \;,\qquad
\label{hprop1} \\[1.ex]
{\bf B}_{i_1} &= &{\bf H} - {\bf C}_{i_1} \; , \qquad 
\label{hprop2} \\[1.ex] 
{\bf B}_{(i_1  \cdots i_m )(j_1j_2 \cdots i_n)} &=& 
{\bf C}_{(i_1 \cdots j_m)} - {\bf C}_{(j_1 j_2 \cdots j_n)}\;\; 
\label{hprop3}
\eeqar{hbgcdef}
the Hard, Cascade, and Bertsch-Gunion propagators 
in the transverse momentum 
space~\cite{Gyulassy:2000er,Gyulassy:2000fs}. In 
Eqs.~(\ref{hprop}), (\ref{hprop1}), (\ref{hprop2}) and (\ref{hprop3}) 
${\bf k}$ 
is the transverse momentum of the emitted gluon and ${\bf q}_{i}$ 
are the transverse momentum transfers from the medium to the 
jet+gluon system at positions $z_i$. Another important quantity, 
which enters the bremsstrahlung spectrum, is the formation time 
of the gluon, $\tau_f$, at the radiation vertex. When compared 
to the separation between the scattering  centers 
$\Delta z_j  = z_j - z_{j-1}$, which can fluctuate up to 
the size of the medium $L$,  it determines the degree of 
coherence present in the multiple scattering process. We introduce 
the following notation~\cite{Gyulassy:2000er,Gyulassy:2000fs}:     
\beqar
\tau_0^{-1} = \omega_0 &=& \frac{ {\bf k}^2 }{k^+}\; , \qquad 
\qquad \label{phases1}  \\[1.ex]
\tau_{i_1 }^{-1} =  \omega_{i_1}
&=&  \frac{ ({\bf k} - {\bf q}_{i_1} )^2  }{k^+} \;, \qquad  
\label{phases2} \\[1.ex]
\tau_{(i_1  \cdots i_m)}^{-1} =  \omega_{(i_1 \cdots i_m)}
&=&  \frac{ ({\bf k} - {\bf q}_{i_1} - 
 \cdots -{\bf q}_{i_m} )^2  }{k^+} \;. \qquad \label{phases3}
\eeqar{phases}
In Eqs. (\ref{phases1}),  (\ref{phases2}) and  (\ref{phases3})
$k^+$ is the large positive lightcone momentum of the 
bremsstrahlung gluon. 
We note that in calculating the amplitudes for the gluon emission,
propagators of the type Eq.~(\ref{hprop}) - Eq.~(\ref{hbgcdef})  
come with a factor $-2ig_s \epsilon_\perp \cdot (\cdots)$ that we 
don't write explicitly, see Appendix~\ref{silmlify}. Similarly, 
at the level of squared amplitudes $\sum_{\rm polarizations} 
\epsilon_\mu^{\perp\, \star}   \epsilon_\nu^{\perp} = 
-g^\perp_{\mu \nu}$. For physical quantities, this leaves us 
with the  product of the 
2D propagators in the amplitude and its conjugate.

\subsection{Constructing the reaction operator}

\begin{figure}[!b]
\begin{center}
\epsfig{file=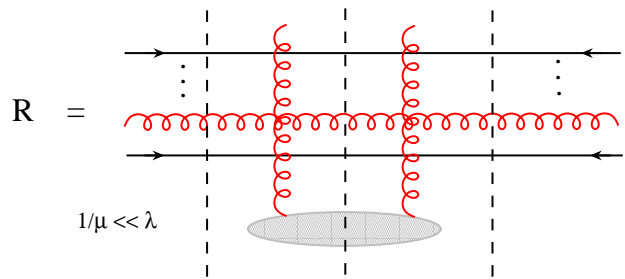,width=3.6in,angle=0}
\vspace*{.1in}
\caption{(Color online) Representation of the reaction operator 
for an arbitrary propagating system of partons. Three
different cuts, corresponding to the relevant 
single-Born and double-Born interactions, are shown. }  
\label{f-rop}
\end{center} 
\end{figure}

In the limit of high energy parton propagation in matter, 
multiple interactions are path ordered. The leading 
nuclear-size-enhanced contribution, $\propto L/\lambda$, to the 
modification of such partonic systems arises from two 
gluon exchanges with the strongly interacting constituents 
of the medium, as shown in Fig.~\ref{f-rop}. Unitarization of 
multiple scattering requires inclusion of three 
distinct cuts in the Feynman diagrams and is also illustrated. 
We note that $n > 2$ gluon exchanges in the region of 
local color neutralization $1/\mu \ll \lambda$ will lead to 
higher order, $\propto \alpha^{n-2}$, corrections to the in-medium 
scattering, which are {\em not} nuclear size-enhanced. These 
are absorbed in the mean free path $\lambda$ for phenomenological 
applications. It is understood that all possible ways of attaching the 
momentum exchanges to the constituent partons of the propagating 
system should be considered, which increases the complexity of 
the calculation with the number of constituents. In the simplest 
case of single parton propagation, one can describe its transverse 
momentum diffusion due to the random walk in nuclear 
matter~\cite{Gyulassy:2002yv}. Recently, the dissociation of
heavy mesons in the QGP has been calculated by applying 
this general approach to a $q\bar{q}$ system~\cite{Adil:2006ra}. 
Finally, the same classes of diagrams, shown in Fig.~\ref{f-rop},
were used to calculate direct photon and dilepton emission to
first order in opacity~\cite{Zhang:2006zd}.

For the purpose of this paper, we are interested in  
a two-parton, jet+gluon, system propagation. The problem
of radiative energy loss in QCD is more complicated
than meson dissociation due to the multiple possibilities
for the location of the gluon emission vertex in 
the many-body scattering.
We first review~\cite{Gyulassy:2000er} the construction 
of the reaction operator,  illustrated  in Fig.~\ref{f-rop}. 
Let ${\cal A}_{i_1\cdots i_n}(x,{\bf k};c)$  be  the 
amplitude when $n$ correlated interactions between the 
system and the medium may have already occurred. 
Here, $x=k^+/p^+$ and ${\bf k}$ are the 
kinematic variables and $c \equiv T_c$ is the color 
matrix associated with the radiative gluon. For each 
scattering center the system may: not interact; 
interact via a single-Born 
scattering; or interact via double-Born scattering. 
These correspond to the three possible cuts when an
amplitude times its conjugate amplitude are considered.
Therefore, the SU(3) color and kinematic modifications 
that arise from in-medium scattering  at position $z_m$ 
are represented by the operators:
\beqar
&& \hat{ 1}_m \; \; -     \qquad   
{\rm Unit \;  operator\; (no \;interaction) }\; ,  
\label{no-i}  \\ 
&& \htD_m \; \; -     \qquad   
{\rm Direct \; operator \; (single-Born)}\; ,  
\label{sing-i} \\
&& \htV_m \; \; -      \qquad   
{\rm Virtual \; operator \; (double-Born)}\; .  \quad
\label{doub-i} 
\eeqar{op-def} 
Thus, starting with an initial condition 
$G_0(x,{\bf k};c)$, which can be the vacuum radiation 
associated with the hard scattering of the parton or the 
lack of such radiation for asymptotic on-shell 
jets~\cite{Gyulassy:2000er,Gyulassy:2000fs},  we can 
construct the amplitude as follows:
\beqar
\vAi(x,{\bf k},c)&=&\prod_{m=1}^n
\left( \delta_{0,i_m} \hat{1} + \delta_{1,i_m} \htD_m 
+ \delta_{2,i_m}  \htV_m \right)   \nonumber \\
 && \times G_0(x,{\bf k};c) \; . 
\eeqar{atens}
In Eq.~(\ref{atens})   $\delta_{i,i_m}$ are the Kronecker
symbols and the indexes $i_m$ keep track of which type of
interaction, Eqs.~(\ref{no-i}), (\ref{sing-i}) and (\ref{doub-i}),  
has occurred.  The operators are ordered from right to 
left as follows: $\hat{ \cal {O}}_{i_n} \hat{ \cal {O}}_{i_{n-1}}
\cdots \hat{ \cal {O}}_{i_{1}}$ for $i_n > i_{n-1} > \cdots > i_1$.
The conjugate amplitude 
$\vAbi(x,{\bf k};c)$ is then uniquely defined since no interaction
is accompanied by a double-Born term, a single-Born interaction 
is accompanied by a single-Born term, and a double-Born interaction 
is accompanied by the unit term, see Fig.~\ref{f-rop}:
\beqar
\vAbi(x,{\bf k},c) & = & G_0^\dagger  (x,{\bf k};c)  
  \times \nonumber \\
&& \hspace*{-1cm}  \prod_{m=1}^n \left( \delta_{0,i_m} 
\hat{V}_m^\dagger + 
\delta_{1,i_m} \hat{D}_m^\dagger 
+ \delta_{2,i_m} \hat{1}^\dagger \right) 
\;  . \qquad
\eeqar{atens2}
In Eq.~(\ref{atens2}) the operators 
are ordered from left to right: 
$  \hat{ \cal {O}}_{i_{1}}^\dagger \cdots 
\hat{ \cal {O}}_{i_{n-1}}^\dagger \hat{ \cal {O}}_{i_{n}}^\dagger$ 
for $i_n > i_{n-1} > \cdots > i_1$ and act to the left.

The contribution to the cross section arising from $n$ correlated
interactions  can then be written as follows:
\beqar
P_n(x,{\bf k}) &= & \vAbi(c)\vAi(c) \nonumber \\
& \equiv&  \tr \sum_{i_1=0}^2\cdots \sum_{i_n=0}^2
{\cal A}^\dagger_{i_1\cdots i_n}(x,{\bf k};c) 
{\cal A}_{i_1\cdots i_n}(x,{\bf k};c)  \;  \quad  \nonumber  \\    
  &=&  \tr \sum_{i_1=0}^2\cdots \sum_{i_{n-1}=0}^2
\bar{\cal A}^{i_1\cdots i_{n-1}}(x,{\bf k};c) \times \nonumber \\ 
&& \left( \hat{D}_n^\dagger
\hat{D}_n+\hat{V}_n+\hat{V}_n^\dagger \right) 
{\cal A}_{i_1\cdots i_{n-1}}(x,{\bf k};c) \; .
\eeqar{pnten}
From Eq.~(\ref{pnten}) we can identify the 
reaction operator,
\beq
\htR \equiv \left( \hat{D}_n^\dagger
\hat{D}_n+\hat{V}_n+\hat{V}_n^\dagger \right) \; , 
\eeq{rop}
as the basic input in this iterative approach to 
multiple scattering.

Clearly, to obtain the contribution from any number, $n$, 
of correlated  interactions to the cross section (or the 
radiative gluon differential distribution) one has to:
\begin{enumerate}
\item Understand the structure of the direct and virtual
operators,  $ \hat{D}_n $ and  $ \hat{V}_n $. 
The unit operator $\hat{1}$ is trivial.
\item Construct the reaction operator, Eq.~(\ref{rop}).
\item Identify the relevant initial condition for the 
problem at hand,  $ \propto  G_0^\dagger  (x,{\bf k};c)  
 G_0  (x,{\bf k};c) $.   
\item Solve the recurrence relation with this initial 
condition.  
\end{enumerate}

\begin{figure}[!tb]
\begin{center}
\epsfig{file=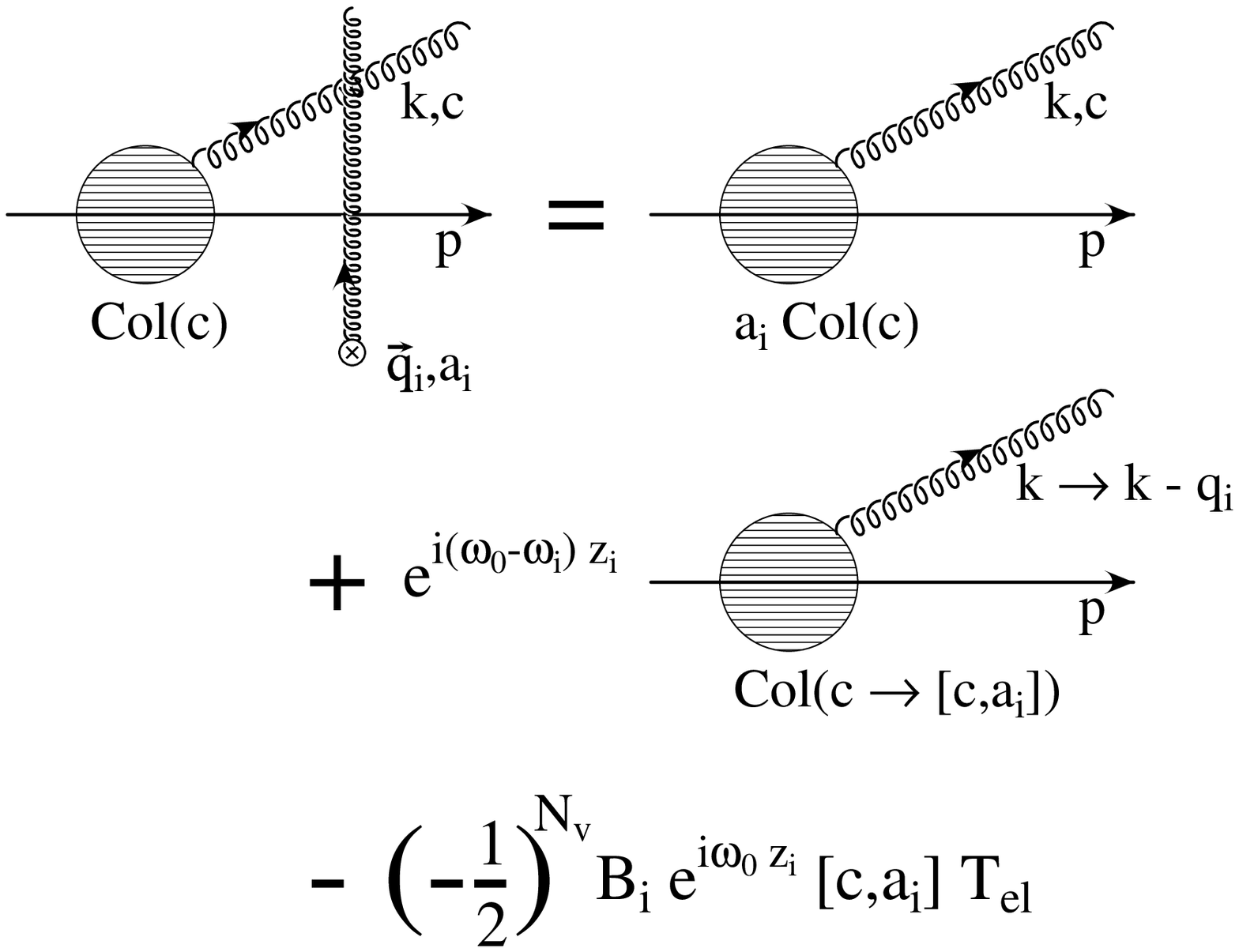,width=3.6in,height=2.6in,angle=0}
\vspace*{0in}
\caption{ Diagrammatic representation of 
the sum of amplitudes generated by the direct single-Born
scattering $ \htD_i {\cal A}$.}
\label{fig-D-schem}
\end{center} 
\end{figure}

We first consider the action of the direct operator 
$\htD_n$ at position $z_n$  on an amplitude with $n-1$ 
correlated scatterings. In this case, there is a single
momentum transfer ${\bf q}_i$ in the amplitude and the 
same momentum transfer due to $\delta^2({\bf q}_i - 
{\bf q}_i^\prime)$  in the conjugate 
amplitude~\cite{Gyulassy:2000er}, see Appendix A.  
The result of such  action on 
${\cal A}_{i_1\cdots i_{n-1}}$  can be represented as:
\beqar
 && \htD_n \vAim(x,{\bf k};c) \equiv    (
a_n \hat{1}+ \htS_n + \htB_n) \vAim(x,{\bf k};c)
 \nonumber \\[1ex]
&&=  a_n \vAim(x,{\bf k};c)   \nonumber \\[1ex]
&& + e^{i(\omega_0-\omega_n)z_n } 
\vAim(x,{\bf k}- {\bf q}_{n};[c,a_n]) \nonumber \\[1ex]
&&  - \left(-\half \,\right )^{N_v(\vAim)} {\bf B}_n \, 
e^{i \omega_0 z_n} [c,a_n] T_{el}(\vAim) \;\;.
\eeqar{didamit}  
The first term in Eq.~(\ref{didamit}) corresponds to a momentum
exchange with the energetic jet. In the high energy limit terms 
of the type $p^- \sim (\sum_i {\bf q}_i)^2/p^+$ are suppressed
and we do not keep track of the transverse modification of the 
parent parton. Such interaction corresponds to an additional
color factor $a_n$. The second term in Eq.~(\ref{didamit})
arises from the momentum transfer ${\bf q}_n$ to the 
radiative gluon, which cannot be neglected since the typical
${\bf k} \sim \sum_i {\bf q}_i $. If the gluon emerges with 
momentum ${\bf k}$ after the momentum transfer, in the 
amplitude  $\vAim $ it has momentum ${\bf k} - {\bf q}_n$.  
The interference phases $e^{i(\omega_0-\omega_n)z_n }$  
arise from the difference in the longitudinal momentum 
components before and after the the gluon interaction. 
Finally, if the gluon is represented by color matrix 
$c$, in the amplitude $\vAim $ color is rotated as follows,
$i f^{c a_n d } d  = [c,a_n]$, i.e. $c \rightarrow [c,a_n]$.

In addition to the above modifications to existing diagrammatic
contributions for the amplitude with $n-1$ scattering centers, 
the acceleration at position $z_n$ acts as a source of a new color 
current, ${\bf B}_n$, with a phase factor $e^{i \omega_0 z_n}$. 
Thus, the last term in Eq.~(\ref{didamit}) represents the 
diagrammatic contributions where the gluon is emitted immediately
before or after the scattering center. In the former case, the 
gluon may also interact with this center. The parent parton has 
a cumulative color factor
\beqar
&& T_{el}({\cal A}_{i_1\cdots i_{n-1}})
\equiv (a_{n-1})^{i_{n-1}}\cdots (a_1)^{i_1}
\;, \qquad \\
&& T_{el}^\dagger(\bar{\cal A}^{i_1\cdots i_{n-1}})
\equiv (a_1)^{2-i_1} \cdots (a_{n-1})^{2-i_{n-1}} \;  . 
\eeqar{tel} 
In the opacity expansion approach there is a unique 
relation between the color factors in the amplitude and
its conjugate: 
\beq
T^\dagger_{el}(\bar{\cal A}^{i_1\cdots i_{n-1}})
T_{el}({\cal A}_{i_1\cdots i_{n-1}}) = C_R^{n-1} \, {\bf 1} \;, 
\eeq{elid}
where $C_R$ is the quadratic Casimir in the fundamental or
adjoint representations for quark or gluon parent partons, 
respectively. We denote by $N_v$ and $\bar{N}_v$ the number
of double-Born interactions in the amplitude and its 
conjugate:  
\beqar
&&N_v=N_v(\vAim)=\sum_{m=1}^{n-1} \delta_{2,i_m}\;,  \qquad 
\\
&&\bar{N}_v=N_v(\vAbim)=\sum_{m=1}^{n-1} \delta_{0,i_m} 
\; ,
\eeqar{kv}
and note that from a multinomial decomposition of zero we have:
\beqar
&&\sum_{i_1,\cdots,i_m} 
\left(-\half\right)^{N_v(\bar{\cal A}_{i_1\cdots i_m})}
\left(-\half\right)^{N_v({\cal A}_{i_1\cdots i_m})}  \nonumber \\ 
&&\hspace*{1.5cm}=\left(-\half -\half +1\right)^m=0 \;  .
\eeqar{kvid}
The factor $(-1/2)^{N_v}$ arises from the symmetry in the 
two gluon legs at the same location $z_m$ and from the fact 
that when both momentum exchanges are in the amplitude or 
its conjugate we have $i^2 = (-i)^2 = -1$.

To summarize, for the medium-induced radiative problem in 
QCD,  the single-Born or direct interaction can be represented 
as follows: 
\beq
 \htD_n  \equiv  ( a_n \hat{1}+ \htS_n + \htB_n) \;. 
\eeq{V-dec}
Here, 
\beq
\htS_n  =i f^{ca_n d} \times e^{i(\omega_0-\omega_n)z_n } 
e^{i{\bf q}_{n} \cdot \hat{\bf b}} \;,
\eeq{S-def}
with  $\hat{\bf b}\equiv i 
\stackrel{\longrightarrow}{{\nabla}_{\bf k}}$ and 
$ e^{i{\bf q}\cdot\hat{\bf b}} 
f({\bf k})=f({\bf k}-{\bf q})$. It is implicit that the
color rotation $if^{ca_nd} d$, yielding $[c,a_n]$, is of the appropriate 
color matrix $d$ not shown explicitly in Eq.~(\ref{S-def}). 
The additional Bertsch-Gunion operator reads: 
\beqar
\htB_n  &=&  - \left(-\half\right)^{N_v(\vAim)}
if^{ca_n d} d  \nonumber \\
&& \times \, {\bf B}_n\,  e^{i \omega_0 z_n}\; T_{el}(\vAim) \;.
\eeqar{B-def}

\begin{figure}[!tb]
\begin{center}
\epsfig{file=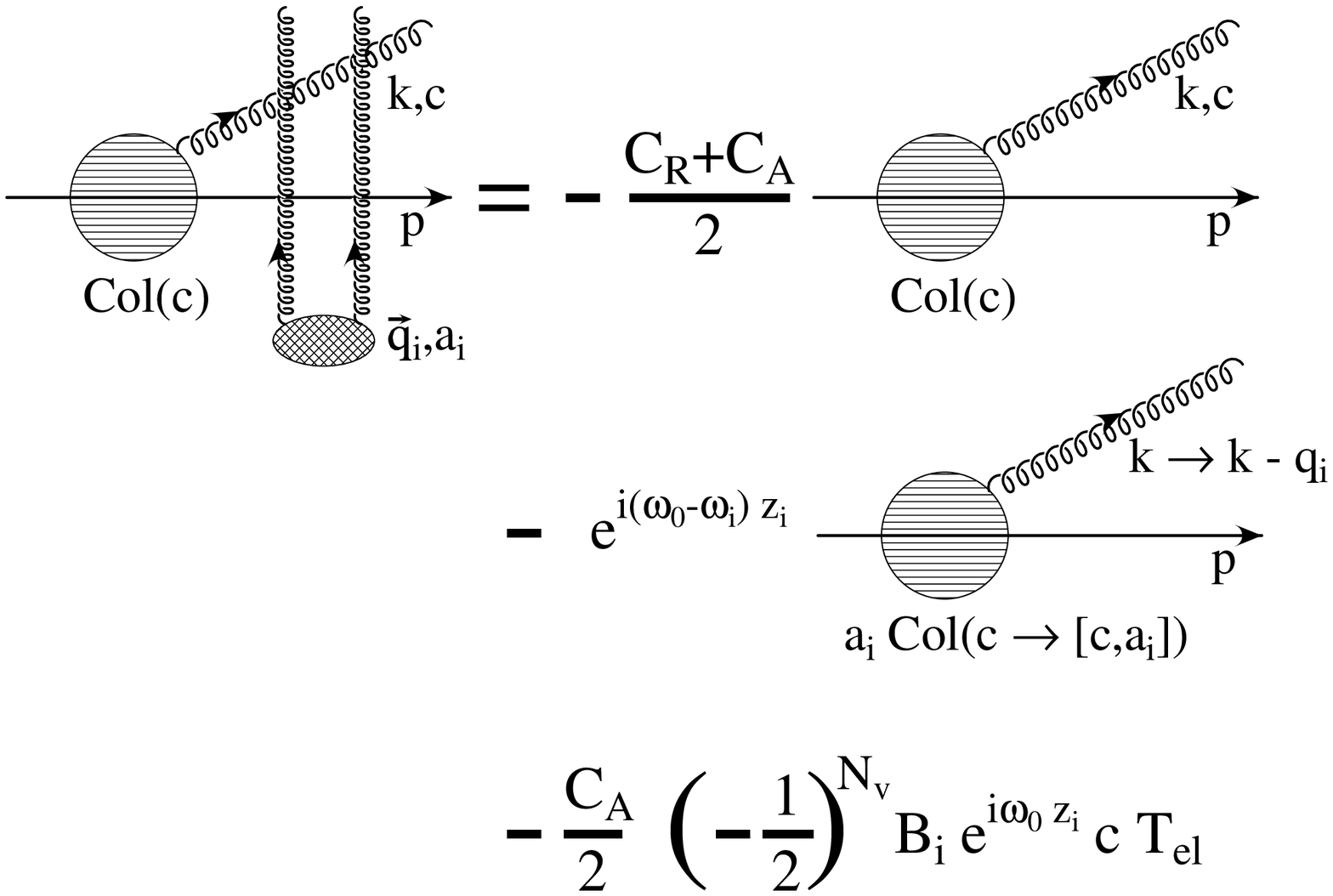,width=3.6in,height=2.6in,angle=0}
\vspace*{.1in}
\caption{   Diagrammatic representation of
the sum of amplitudes  generated by the virtual double-Born
scattering  $ \htV_i {\cal A}$. }
\label{fig-Doub}
\end{center} 
\end{figure}

Next, we consider the double-Born interaction of the jet
at position $z_n$, as shown in Fig.~\ref{fig-Doub}.  
In this case there are two equal and opposite momentum 
transfers ${\bf q}_i$ in the amplitude (or the conjugate
amplitude) due to $\delta^2({\bf q}_i + 
{\bf q}_i^\prime)$~\cite{Gyulassy:2000er}, 
see Appendix A. The modification of the amplitude  
$\vAim(x,{\bf k};c)$ is found to be:
\beqar
&& \htV_n \vAim(x,{\bf k};c)
=  - \frac{C_R+C_A}{2} \vAim(x,{\bf k};c) \nonumber \\[1ex] 
&& -  e^{i(\omega_0-\omega_n)z_n } a_n
\vAim(x,{\bf k}- {\bf q}_{n}, [c,a_n]) \nonumber \\[1ex]
&& - \left(-\half\,\right)^{N_v(\vAim)}
\frac{C_A}{2} \, {\bf B}_n \,
e^{i \omega_0 z_n} c  T_{el}(\vAim)     \;. \qquad
\eeqar{vidamit}

The first term in Eq.~(\ref{vidamit}) corresponds to the 
case when both momentum exchanges are with the parent parton 
or the radiative gluon. In this case there is no net 
momentum transfer, no additional phase factors arise and
the two color matrices yield the quadratic Casimirs $C_R$ 
and  $C_A$. The factor $(-1/2)$ was discussed above.
The second term in  Eq.~(\ref{vidamit}) represents the 
situation where one of the gluon legs is attached to the 
jet and one to the bremsstrahlung gluon. Since there are
two possible combinations $2 \times (-1/2) = -1$ . The 
shift in the transverse momentum ${\bf k} \rightarrow 
{\bf k} - {\bf q}_n$, the phase factor  
$e^{i(\omega_0-\omega_n)z_n }$ and the color rotation 
$i f^{ca_n d} d$, i.e. $c \rightarrow [c,a_n]$, are all the same 
as in the second term in Eq.~(\ref{didamit}). The difference 
is that the interaction with the parent parton gives an 
additional color factor $a_n$. Finally, parts of the 
diagrams where a gluon is emitted immediately before or 
after the interaction point $z_n$ are combined in the last
term in Eq.~(\ref{vidamit}). This term is identical to 
the last term in Eq.~(\ref{didamit}), except for the 
color factor since we have carried out the simplification 
$a_n [c,a_n] =  - (C_A/2) c$. The additional $(-1)$ arises from 
the last two gluon exchanges at $z_n$.  
In summary, the double-Born interaction at $z_n$  can 
be implemented by the following operator
\beqar
\htV_n  &=&  -\half(C_A+C_R)\hat{1} - a_n 
\htS_n- a_n\htB_n \nonumber \\
 &=&  -a_n \htD_n - \half(C_A-C_R)  \;  ,
\eeqar{dvid} 
and is clearly not linearly independent of the single-Born
term, Eq.~(\ref{V-dec}).

Now, we can proceed and construct the reaction operator $\htR_n$   
to relate the  $n^{\rm th}$  order in opacity  gluon emission 
``probability'' distribution  $P_n(x,k)$ to the probability at 
$(n-1)^{\rm st}$ order.
We express the reaction operator in Eq.~(\ref{rop}) as
follows:
\beqar
\htR_n &=&  (\htD_{n}-a_{n})^\dagger  
(\htD_{n}-a_{n}) - C_A \hat{1} \nonumber \\[1ex]
&=& 
(\htS_{n}+\htB_{n})^\dagger  (\htS_{n}+\htB_{n}) 
-C_A \hat{1} \nonumber \\[1ex]
&=&  (\htS_{n}^\dagger \htS_{n} - C_A 
\hat{1}) + \htB_{n}^\dagger 
\htB_{n} + 2 {\cal R}e \; \htS_{n}^\dagger 
\htB_{n} \; . \quad
\eeqar{np1} 
Noting that $if^{ca_{n} d} (-i)f^{ca_{n} d^\prime} = 
C_A \delta_{d,d^\prime}$, we see that the first term  
in Eq.~(\ref{np1}),
\beqar
 (\htS_{n}^\dagger \htS_{n} - C_A 
\hat{1}) &=&  
C_A (e^{-{\bf q}_{n} \stackrel{\leftarrow}{{\nabla}_{{\bf k}}}}
e^{-{\bf q}_{n} \stackrel{\rightarrow}{{\nabla}_{{\bf k}}  } } 
- \hat{1} )  \;,  
\eeqar{s-simp}
gives a homogeneous contribution to the functional 
recurrence relation: 
\beqar
&&  \vAbim(x,{\bf k};c) (\htS_{n}^\dagger \htS_{n} - C_A )
\vAim(x,{\bf k};c )     \nonumber \\[1ex] 
&&  \qquad  = C_A \left(  P_{n-1}(x,{\bf k}-{\bf q}_{n} )
            - P_{n-1}(x,{\bf k}) \right)  \nonumber \\[1ex] 
& & \qquad = C_A \left( e^{i{\bf q}_{n}\cdot\hat{\bf b}}-1 \right) 
P_{n-1}(x,{\bf k}) \; .
\eeqar{assa}
The second term does not contribute beyond first order in 
opacity, $n=1$, since 
\beqar
&& 
\vAbim(x,{\bf k};c)\htB_{n}^\dagger \htB_{n}\vAim(x,{\bf k};c) 
= C_A \, |{\bf B}_n|^2  \nonumber \\[1ex]
&& \quad \times
\sum_{i_1,\cdots,i_{n-1}}  \left(-\half\right)^{\bar{N}_v(\vAbim)}
\left(-\half\right)^{N_v(\vAim)} \nonumber \\
&& \quad 
\times \; T_{el}^\dagger(\vAbim) \, c c \, T_{el}(\vAim) 
 \equiv  0 \; .  \qquad
\eeqar{bb}
For this proof we used Eq.~(\ref{elid}), and the  identity~(\ref{kvid}).
For $n=1$, however, the $C_R C_A |{\bf B}_1|^2$ contribution
to  $P_{n-1}(x,{\bf k})$ survives.
Finally, the non-diagonal term in Eq.~(\ref{np1}) reads:
\beqar
\hspace*{-.5cm}&& 2 {\cal R}e \, \vAbim(x,{\bf k};c) \htB_{n}^\dagger 
\htS_{n}\vAim(x,{\bf k};c) \nonumber \\[1ex] 
& & = -2 C_A\, {\bf B}_n\cdot \left({\rm Re}\; e^{-i\omega_nz_n} 
e^{i{\bf q_n}\cdot\hat{\bf b}} {\bf I}_{n-1}\right)
\; , 
\eeqar{bcn}
where we have used Eqs.~(\ref{S-def}) and (\ref{B-def}).  
Writing explicitly the expression for ${\bf I}_{n-1}$ 
and representing the $(n-1)^{\rm th}$ step in this iteration via
the $\htS_{n-1}$ and $\htB_{n-1}$ operators 
we obtain: 
\beqar
&& {\bf I}_{n-1}  =   \sum_{i_1\cdots i_{n-1}}
\left(-\half\right)^{N_v(\bar{\cal A}^{i_1\cdots i_{n-1}})}
 T_{el}^\dagger(\bar{\cal A}^{i_1\cdots i_{n-1}}) 
\nonumber \\[1ex]
&& \hspace*{1.5cm}\times  c {\cal A}_{i_1\cdots i_{n-1}} (x,{\bf k};c)  
\nonumber \\[.5ex]
&& =  \sum_{i_1\cdots i_{n-2}} 
\left(-\half\right)^{N_v(\bar{\cal A}_{i_1\cdots i_{n-2}})}
 T_{el}^\dagger(\bar{\cal A}_{i_1\cdots i_{n-2}}) 
\nonumber \\[.5ex]
&&  \times \, \left[ C_A(e^{i(\omega_0-\omega_{n-1})z_{n-1}}
 e^{i{\bf q}_{n-1}\cdot\hat{\bf b}} -1) c
+[a_{n-1},c]\htB_{n-1} \right] \nonumber \\[1ex]
&& \times \, {\cal A}_{i_1\cdots i_{n-2}}  (x,{\bf k};c)  \; .
\eeqar{bss2}
In Eq.~(\ref{bss2}) the term proportional to 
$\htB_{n-1}  {\cal A}_{i_1\cdots i_{n-2}}$ vanishes for 
$n > 2$. The proof,  up to an insignificant difference in color 
factors, is again based on the multinomial decomposition of 
zero, Eq.~(\ref{kvid}).

From Eqs.~(\ref{s-simp}) and (\ref{bcn}), the basic iteration 
step for the full solution of the problem of medium induced 
gluon radiation  can be written as follows: 
%\begin{widetext}
\beqar
 P_{n}(x,{\bf k})&=& C_A
\left( P_{n-1}(x,{\bf k}-{\bf q}_{n}) - P_{n-1} (x, {\bf k}) \right)
\nonumber \\[1.ex] 
&& -2 C_A\, {\bf B}_n\cdot \left( {\cal R}e \; e^{-i\omega_nz_n} 
e^{i{\bf q_n}\cdot\hat{\bf b}} {\bf I}_{n-1} \right)
\nonumber \\[1.ex] 
&& + \delta_{n,1} C_A C_R |{\bf B}_1|^2  
\; .  
\eeqar{iter}
The inhomogeneous term in  Eq.~(\ref{iter}) is  expressed as:
\beqar
{\bf I }_{n-1}  & = & C_A 
\left( e^{i(\omega_0-\omega_{n-1})z_{n-1} }
e^{i{\bf q}_{n-1}\cdot\hat{\bf b}} -1 \right) {\bf I}_{n-2}
\nonumber \\[1.ex] 
&&-\delta_{n-1,1} C_A C_R {\bf B}_{1} e^{i\omega_0 z_{1}}
\, , 
\eeqar{bfin}
%\end{widetext}
which is a direct consequence of Eq.~(\ref{bss2}).

We emphasize that each radiative problem yields a 
{\em different} solution, related to the boundary condition
$G_0(x,{\bf k};c)$. For final-state radiation this
initial condition is given by the bremsstrahlung 
associated with the hard scattering of incoming 
partons. This case has been considered in detail 
in Ref.~\cite{Gyulassy:2000er} as the first complete 
application of the GLV approach. Explicit solution 
to all orders in the correlation between the momentum 
transfers from the multiple scattering
centers to the jet+gluon system, suitable for further 
analytic approximations or numerical 
simulations, has been found for the case of final-state radiation,
the third case illustrated in Fig.~\ref{fig-schem}. We emphasize 
that these are not correlations between the scattering centers
of the target arising from the strong nuclear force.   
Subject to the coherence criterion, $\tau_f \geq
\max ( |z_i - z_j|, 0 \leq i,j \leq n ) $  at 
$n^{\rm th}$ order in opacity, the 
momentum transfers are correlated in the sense that the 
bremsstrahlung from the hard collision (at position $z_0$) 
and the multiple soft interactions (at positions $z_j > z_0 $) 
interfere to give a contribution 
to the medium-induced spectrum of gluons.  For completeness,
we quote the result here~\cite{Gyulassy:2000er}:  
\begin{widetext}
\beqar
k^+ \frac{dN^g(FS)}{dk^+ d^2 {\bf k} } &=& \frac{C_R \alpha_s}{\pi^2}
\sum_{n=1}^{\infty}  \left[ \prod_{i = 1}^n  \int
\frac{d \Delta z_i}{\lambda_g(z_i)}  \right] 
\left[ \prod_{j=1}^n \int d^2 {\bf q}_j \left( \frac{1}{\sigma_{el}(z_j)} 
\frac{d \sigma_{el}(z_j) }{d^2 {\bf q}_j}   
-  \delta^2 ({\bf q}_j) \right)    \right] \nonumber \\ 
&& \times \;  \left[ -2\,{\bf C}_{(1, \cdots ,n)} \cdot 
\sum_{m=1}^n {\bf B}_{(m+1, \cdots ,n)(m, \cdots, n)} 
\left( \cos \left (
\, \sum_{k=2}^m \omega_{(k,\cdots,n)} \Delta z_k \right)
-   \cos \left (\, \sum_{k=1}^m \omega_{(k,\cdots,n)} 
\Delta z_k \right) \right)\; \right]  \;,   \qquad 
\eeqar{full-final}
where $\sum_2^1 \equiv 0$ and  ${\bf B}_{(n+1, n)} \equiv {\bf B}_n$ 
is understood. In the case of final-state interactions, 
$z_0 \approx 0$ is the point of the initial hard scatter and $z_L = L$ is 
the extent of the medium.  The path ordering  of the 
interaction points, $z_L > z_{j+1} > z_j > z_0$, leads 
to the  constraint 
$\sum_{i=1}^n \Delta z_i  \leq  z_L $. One implementation of this
condition would be $\Delta z_i \in [\, 0,z_L -\sum_{j=1}^{i-1} 
\Delta z_j \, ]$ 
and it is implicit in Eq.~(\ref{full-final}).
If we write the 2D propagators and interference phases  
in Eq.~(\ref{full-final}) more explicitly,  we have: 
\beqar
k^+ \frac{dN^g(FS)}{dk^+ d^2 {\bf k} } &=& \frac{C_R \alpha_s}{\pi^2}
 \sum_{n=1}^{\infty} \left[ \prod_{i = 1}^n \int
{d \Delta z_i} \; { \sigma_{el}(z_i) \rho (z_i)}  \right] 
\left[ \prod_{j=1}^n \int d^2 {\bf q}_j    
 \left( \frac{1}{\sigma_{el}(z_j)} 
\frac{d \sigma_{el}(z_j) }{d^2 {\bf q}_j}   
-  \delta^2 ({\bf q}_j) \right)    \right]  \nonumber \\ 
&& \hspace*{-0.5cm} \times  \; 
\left[  -2 \;  \frac{{\bf k} - {\bf q}_1 - \cdots - {\bf q}_n}
{({\bf k} - {\bf q}_1 - \cdots - {\bf q}_n)^2} 
\cdot  \sum_{m=1}^n 
   \left( \frac{{\bf k} - {\bf q}_{m+1} - \cdots - {\bf q}_n}
{({\bf k} - {\bf q}_{m+1} - \cdots - {\bf q}_n)^2}  - 
 \frac{{\bf k} - {\bf q}_m - \cdots - {\bf q}_n}
{({\bf k} - {\bf q}_m - \cdots - {\bf q}_n)^2}     \right) 
 \right.  \nonumber \\
&& \left. \times \; \left(  \cos \left( \sum_{k=2}^m 
\frac{ ({\bf k} - {\bf q}_k - \cdots - {\bf q}_n)^2 }{k^+} \Delta z_j   
 \right)   -  \cos \left( \sum_{k=1}^m  
\frac{ ({\bf k} - {\bf q}_k - \cdots - {\bf q}_n)^2 }{k^+} \Delta z_j   
\right) \; \right) \;    \right] \;.
\eeqar{finalfull-det}

\end{widetext}
Jet quenching in the QGP, the best known application of non-Abelian 
energy loss in heavy ion reactions, is based on such a solution. 
For details, see~\cite{Gyulassy:2003mc}.

\subsection{Bertsch-Gunion radiation to all orders in opacity}

The first {\em new} solution that we obtain is this paper is for the
case of asymptotic on-shell jets, initially considered by Bertsch
and Gunion~\cite{Gunion:1981qs} and illustrated as the first 
case in Fig.~\ref{fig-schem}. Although it is not directly 
applicable to the physics situation of high-$p_T$ particle production
due to the lack of hard scattering, it is a necessary step
to fully solve the problem of initial-state energy loss.    
The absence  of hard bremsstrahlung  yields a simple 
initial condition:
\beq
G_0(x,{\bf k};c) = 0  \;. 
\eeq{BG-init}
The solution for the ${\bf I }_{n-1}$  part in the 
inhomogeneous term of Eq.~(\ref{iter}) reads:
\beqar
{\bf I }_{0} &=& 0 \;,  \\[1ex]
{\bf I }_{1} &=& - C_A C_R {\bf B}_{1} e^{i\omega_0 z_{1}} \;,  \\[1ex]
{\bf I }_{n-1} &=& - C_R C_A^{n-1} 
\prod_{m=2}^{n-1}  \left( e^{i(\omega_0-\omega_{m})z_{m} }
 e^{i{\bf q}_{m}\cdot\hat{\bf b}} -1 \right) \nonumber \\[1ex]
&&\times \, {\bf B}_{1} e^{i\omega_0 z_{1}} \; . 
\eeqar{BG-initI}  
Consequently, the solution for $P_n(x,{\bf k})$  can be written 
in the  form 
\begin{widetext}
\beqar 
P_n(x,{\bf k}) &=&  C_R C_A^n \prod_{m=2}^{n}  
(e^{i{\bf q}_{m} \cdot\hat{\bf b}} -1) 
|{\bf B}_{1}|^2 + 2 {\cal R}e \,  C_R C_A^n  \sum_{m=2}^n 
\left[  \prod_{j=m+1}^{n}  (e^{i{\bf q}_{j}\cdot\hat{\bf b}} -1)  
\right]  \, {\bf B}_{m} e^{i{\bf q}_{m} \cdot\hat{\bf b}} 
 e^{-i \omega_0 z_{m} } \nonumber \\[1ex]
&& \times \; 
\prod_{j=2}^{m-1}   (e^{i(\omega_0-\omega_{j})z_{j} }
 e^{i{\bf q}_{j}\cdot\hat{\bf b}} -1)  
\, {\bf B}_{1} e^{i\omega_0 z_{1}} \; . 
\eeqar{pn-bg}
In Eq.~(\ref{pn-bg}) we recall that, for the special case 
of $n=1$, $\sum_{i=2}^1 = 0$ and 
$\prod_{i=2}^1 = 1$.  It is easy to verify that our result is a 
solution of the master recurrence relation  Eq.~(\ref{iter}) by 
rewriting it in the form: 
\beqar 
P_n(x,{\bf k}) & = & C_A (e^{i{\bf q}_{n} \cdot\hat{\bf b}} -1) 
\left[  C_R C_A^{n-1} \prod_{m=2}^{n-1}  
(e^{i{\bf q}_{m} \cdot\hat{\bf b}} -1) 
|{\bf B}_{1}|^2 + 2 {\cal R}e \,  C_R C_A^{n-1}  \sum_{m=2}^{n-1} 
\left[  \prod_{j=m+1}^{n-1}  (e^{i{\bf q}_{j}\cdot\hat{\bf b}} -1)  
\right]  \, {\bf B}_{m} e^{i{\bf q}_{m} \cdot\hat{\bf b}} 
 e^{-i \omega_0 z_{m} } \right.  \nonumber \\[1ex]
&& \left. \times \; 
\prod_{j=2}^{m-1}  C_A (e^{i(\omega_0-\omega_{j})z_{j} }
 e^{i{\bf q}_{j}\cdot\hat{\bf b}} -1)  
\, {\bf B}_{1} e^{i\omega_0 z_{1}} \right]  \;
+  2 {\cal R}e \,  C_R C_A^{n}    
{\bf B}_{n} e^{i{\bf q}_{n} \cdot\hat{\bf b}} 
 e^{-i \omega_0 z_{n} }  \prod_{j=2}^{n-1}  C_A 
(e^{i(\omega_0-\omega_{j})z_{j} } e^{i{\bf q}_{j}\cdot\hat{\bf b}} -1)  
 \nonumber \\[1ex]
&&  \times \; {\bf B}_{1} e^{i\omega_0 z_{1} } \; . 
\eeqar{bgavui}

\end{widetext}

To obtain the contribution of $n$ correlated scatterings to
medium-induced gluon production we have to average over the 
momentum transfers ${\bf q}_n$ in Eq.~(\ref{pn-bg}). Let
$(1/\sigma_{el}) d \sigma_{el}/d^2 {\bf q}_m$ be the differential 
distribution of momentum transfers at position $z_m$. We note 
that the first term in Eq.~(\ref{pn-bg}) can be written as  
follows,
\beqar
&& \left[ \prod_{i=1}^n  \int d^2 {\bf q}_i \; 
\frac{1}{\sigma_{el}}
\frac{d\sigma_{el}}{d^2 {\bf q}_i } \right]
\prod_{m=2}^{n} ( e^{i{\bf q}_{m} \cdot\hat{\bf b}} - 1 ) 
|{\bf B}_{1}|^2  \nonumber \\[.5ex]
& =& \left[ \prod_{i=1}^n  \int d^2 {\bf q}_i \; 
\left( \frac{1}{\sigma_{el}} \frac{d\sigma_{el}}{d^2 {\bf q}_i } 
- \delta^2({\bf q}_i ) \right) \right] \,   \nonumber \\[.5ex]
&&  \times \,  \prod_{m=2}^{n}  e^{i{\bf q}_{m} 
\cdot\hat{\bf b}} \; |{\bf B}_{1}|^2  \;. 
\eeqar{momav}
In Eq.~(\ref{momav}) we are able to carry out the simplification
including the $n=1$ term since 
\beq
\delta^2({\bf q}_i) \,  {\bf B}_{i} = 0  \; .
\eeq{deltaid}
The inhomogeneous term in Eq.~(\ref{pn-bg}) can also be simplified. 
For the momentum transfers ${\bf q}_{m+1}, \cdots, {\bf q}_{n}$ the 
result follows from Eq.~(\ref{momav}).  For the momentum 
transfers ${\bf q}_{2}, \cdots, {\bf q}_{m-1}$ we use 
\begin{eqnarray}
&&  \left[  \prod_{i=2}^{m-1}  \int d^2 {\bf q}_j \; 
\frac{1}{\sigma_{el}} 
\frac{d \sigma_{el}}{d^2 {\bf q}_i } \right] 
\prod_{j=2}^{m-1}  \left( e^{i(\omega_0-\omega_{j})z_{j} } 
e^{i{\bf q}_{j}\cdot\hat{\bf b}} - 1 \right)
\nonumber \\[.5ex]
&&  \times {\bf B}_{1} e^{i\omega_0 z_{1}} \nonumber \\[.5ex] 
&=& \left[ \prod_{i=2}^{m-1}  \int d^2 {\bf q}_i \; 
\left( \frac{1}{\sigma_{el}} \frac{d\sigma_{el}}{d^2 {\bf q}_i } 
- \delta^2({\bf q}_i )  \right)  \right]   \nonumber  \\
&& \times  \prod_{j=2}^{m-1}  e^{i(\omega_0-\omega_{j})z_{j} } 
e^{i{\bf q}_{j}\cdot\hat{\bf b}}  \;  
{\bf B}_{1} e^{i\omega_0 z_{1}} \; .   
\label{mod111}
\\
&& \nonumber
\label{mod1}
\end{eqnarray}
We note that the terms $n=1$ and $n=m$  can also be included 
in the general representation, Eq.~(\ref{mod111}),  
due to Eq.~(\ref{deltaid}). Thus, we write the contribution 
of the multiple scattering centers as follows:
\begin{widetext}
\beqar
 && \left[ \prod_{i=1}^n  \int d^2 {\bf q}_i \; 
\frac{1}{\sigma_{el}}
\frac{d\sigma_{el}}{d^2 {\bf q}_i } \right] \;  
P_n(x,{\bf k})  =  
C_R C_A^n  \left[ \prod_{i=1}^n  \int d^2 {\bf q}_i \; 
 \left( \frac{1}{\sigma_{el}}
\frac{d\sigma_{el}}{d^2 {\bf q}_i } 
- \delta^2({\bf q}_i ) \right) \right] 
\; \left[  \prod_{m=2}^{n}  e^{i{\bf q}_{m} \cdot\hat{\bf b}} \;
|{\bf B}_{1}|^2 \right.  \nonumber \\[1ex] 
&& \qquad \qquad \qquad  \left.  + 2 {\cal R}e \,  \sum_{m=2}^n 
\left[  \prod_{j=m+1}^{n}  e^{i{\bf q}_{j}\cdot\hat{\bf b}}   
\right]  \, {\bf B}_{m} e^{i{\bf q}_{m} \cdot\hat{\bf b}} 
 e^{-i \omega_0 z_{m} } \; 
\prod_{j=2}^{m-1} e^{i(\omega_0-\omega_{j})z_{j} }
 e^{i{\bf q}_{j}\cdot\hat{\bf b}}   
\, {\bf B}_{1} e^{i\omega_0 z_{1}}  \right]\; . 
\eeqar{rep2}

\end{widetext}
We note again that in Eq.~(\ref{rep2}) the products of momentum 
shift  operators  or phases and momentum shift operators are 
applied sequentially from the left with each operator acting on the 
function resulting from the previous one.

The last step in writing the explicit solution is to 
carry out the action of the momentum shift operators in 
Eq.~(\ref{rep2}). The first homogeneous term can be easily 
simplified  since  
\beqar
\prod_{m=2}^{n} e^{i{\bf q}_{m} \cdot\hat{\bf b}} \;
|{\bf B}_{1}|^2  =  |{\bf B}_{(2\cdots n)(1 \cdots n)}|^2 \;.
\eeqar{bgsimp1}
The inhomogeneous term can also be simplified as 
follows: 
\begin{eqnarray} 
\nonumber
&&  \prod_{j=m+1}^{n}  e^{i{\bf q}_{j}\cdot\hat{\bf b}}   
 \, {\bf B}_{m} e^{i{\bf q}_{m} \cdot\hat{\bf b}} 
 e^{-i \omega_0 z_{m} }  \;  \nonumber \\[1ex]
&& \qquad \qquad \times  \prod_{j=2}^{m-1} e^{i(\omega_0-\omega_{j})z_{j} }
 e^{i{\bf q}_{j}\cdot\hat{\bf b}}   
\, {\bf B}_{1} e^{i\omega_0 z_{1}}  \nonumber \\[1ex]
&&   = {\bf B}_{(m+1\cdots n)(m \cdots n)} 
e^{i \sum_{j=m}^n {\bf q}_{j}\cdot\hat{\bf b} } \nonumber \\[1ex]
&& \;\;  \times  \, e^{-i \sum_{j=2}^m \omega_{(j \cdots m-1)} 
(z_j - z_{j-1})} {\bf B}_{(2\cdots m-1)(1 \cdots m-1)} \; . \quad
\nonumber \\[1ex]
&& = {\bf B}_{(m+1\cdots n)(m \cdots n)} \cdot  
{\bf B}_{(2\cdots n)(1 \cdots n)} 
\nonumber \\[1ex]
&& \quad 
\times e^{-i \sum_{j=2}^m \omega_{(j \cdots m-1)} \Delta z_j } \; . 
%\nonumber \\
\label{bgsimp2} 
\end{eqnarray}

The overall normalization for the differential gluon distribution 
is set by the
color factor, the strong coupling constant and a phase space 
factor that combine to produce the factor $C_R \alpha_s/\pi^2$. 
We note that $C_R$ signifies the color charge dependence 
of the gluon bremsstrahlung 
in the small energy loss limit $\Delta E_{\rm rad} / E \ll 1$. 
Similarly to the case of final state gluon bremsstrahlung, in the
limit of small lightcone momentum fractions $x$ and small 
transverse momenta  ${\bf k}$  the result is ``color trivial'',
retaining  only the quadratic Casimirs in the adjoint 
representation. These enter the mean free 
path of the propagating jet+gluon system, indicating that only 
gluon rescattering is important, i.e. $\lambda_g(z_i)$. Putting 
everything together we find:
\begin{widetext} 
\beqar
k^+ \frac{dN^g(BG)}{dk^+ d^2 {\bf k} } &=& \frac{C_R \alpha_s}{\pi^2}
\left[ \prod_{i = 1}^n \int
\frac{d \Delta z_i}{\lambda_g(z_i)}  \right] 
\left[ \prod_{j=1}^n \int d^2 {\bf q}_j \left( \frac{1}{\sigma_{el}(z_j)} 
\frac{d \sigma_{el}(z_j) }{d^2 {\bf q}_j}   
-  \delta^2 ({\bf q}_j) \right)    \right] \nonumber \\ 
&& \times  \;  {\bf B}_{(2\cdots n)(1\cdots n)} \cdot 
\left[ {\bf B}_{(2\cdots n)(1\cdots n)} 
 + 2 \sum_{i=2}^n {\bf B}_{(i+1 \cdots n)(i\cdots n)} 
\cos \left( \sum_{j=2}^i \omega_{(j\cdots n)} \Delta z_j  
\right) \right] \;.
\eeqar{BGfull}
A direct comparison of this general result to the brute force calculation 
up to second order in opacity can be found in Appendix~\ref{bg2ord}.
The integration limits on the separation between the multiple scattering 
centers were discussed in the previous section. Since in this case there
is no hard scattering, the maximum separation corresponds to the physical 
size of the  medium. More explicitly, our result reads:
\beqar
k^+ \frac{dN^g(BG)}{dk^+ d^2 {\bf k} } &=& \frac{C_R \alpha_s}{\pi^2}
\left[ \prod_{i = 1}^n \int
{d \Delta z_i}\; { \sigma_{el}(z_i) \rho(z_i)}  \right] 
\left[ \prod_{j=1}^n \int d^2 {\bf q}_j    
 \left( \frac{1}{\sigma_{el}(z_j)} 
\frac{d \sigma_{el}(z_j) }{d^2 {\bf q}_j}   
-  \delta^2 ({\bf q}_j) \right)    \right]  \nonumber \\ 
&& \hspace*{-0.5cm} \times  \; 
\left( \frac{{\bf k} - {\bf q}_2 - \cdots - {\bf q}_n}
{({\bf k} - {\bf q}_2 - \cdots - {\bf q}_n)^2}  - 
 \frac{{\bf k} - {\bf q}_1 - \cdots - {\bf q}_n}
{({\bf k} - {\bf q}_1 - \cdots - {\bf q}_n)^2}     \right)
\cdot 
\left[ \left( \frac{{\bf k} - {\bf q}_2 - \cdots - {\bf q}_n}
{({\bf k} - {\bf q}_2 - \cdots - {\bf q}_n)^2}  - 
 \frac{{\bf k} - {\bf q}_1 - \cdots - {\bf q}_n}
{({\bf k} - {\bf q}_1 - \cdots - {\bf q}_n)^2}     \right) 
 \right. \nonumber \\
&& \hspace*{-0.5cm} \left.  + 2 \sum_{i=2}^n 
\left( \frac{ {\bf k} - {\bf q}_{i+1} - \cdots - {\bf q}_n }
{ ({\bf k} - {\bf q}_{i+1} - \cdots - {\bf q}_n )^2 }  - 
 \frac{ {\bf k} - {\bf q}_i - \cdots - {\bf q}_n }
{ ( {\bf k} - {\bf q}_i - \cdots - {\bf q}_n)^2 }     \right) 
\cos \left(  \sum_{j=2}^i  
\frac{ ({\bf k} - {\bf q}_j - \cdots - {\bf q}_n)^2 }{k^+} \Delta z_j   
 \right) \right] \;.
\eeqar{BGfull-det}

\end{widetext}
We recall the convention ${\bf B}_{(n+1, n)} \equiv {\bf B}_n$. Thus, 
to first order in opacity the Bertsch-Gunion case of asymptotic 
$t=-\infty$ to $t=+\infty$ on-shell jets yields a gluon radiative spectrum
$\propto |{\bf B}_1 |^2 $ and no coherence effects. Note that in 
Eq.~(\ref{BGfull-det}) $\sigma_{el}$ is the gluon scattering cross 
section.

\subsection{Initial state radiation to all orders in opacity}

For hadronic reactions where high-$p_T$/high-$E_T$ particles/jets 
are detected, the relevant initial-state interactions 
that lead to energy loss are illustrated as the second 
case in Fig.~(\ref{fig-schem}).
The difference from the Bertsch-Gunion case is that 
there is always radiation associated with the hard 
scatter at position $z_L$. In particular, it can be
written as a boundary contribution in the absence of
soft momentum transfers from the medium in the form:
\beq
H(x,{\bf k};c)  
= + {\bf{H}} e^{i \omega_0 z_L} \;. 
\eeq{H-init}
Such a term will always be present in the opacity expansion
of the amplitude and its conjugate but modulated by 
the color and symmetry factors associated with the 
preceding interactions of the parent jet with 
the nuclear matter. More specifically, 
\beqar
&& \vAi (x,{\bf k};c)_{IS} =  \vAi (x,{\bf k};c)_{BG}  \nonumber \\[1ex]
&&  + \left(-\frac{1}{2} \right)^{ {N}_v (\vAi) }
c T_{el} ( {\cal A}_{i_1 \cdots i_n}) \, {\bf{H}} e^{ i \omega_0 z_L } 
\; , \\[2ex]
&& \vAbi (x,{\bf k};c)_{IS} = \vAbi (x,{\bf k};c)_{BG} \nonumber \\[1ex]
&& +  \left( -\frac{1}{2} \right)^{ \bar{N}_v(\vAbi) } 
T_{el}^\dagger (\vAbi) 
{ \bf H } e^{-i \omega_0 z_L} \; .
\eeqar{111}
The contribution to the differential distribution at $n^{\rm th}$ 
order in opacity is proportional to
\beqar
&&P_n(x,{\bf k})_{IS} = \vAbi(x,{\bf k};c)_{IS} \vAi(x,{\bf k};c)_{IS} 
\nonumber \\[1ex] 
&& \equiv  \tr \sum_{i_1=0}^2\cdots \sum_{i_n=0}^2
\bar{\cal A}^\dagger_{i_1\cdots i_n}(x,{\bf k};c)_{IS} 
{\cal A}_{i_1\cdots i_n}(x,{\bf k};c)_{IS}  \; \quad  \nonumber  \\[1ex]     
&& =   \vAbi(x,{\bf k};c)_{BG} \vAi(x,{\bf k};c)_{BG} 
\quad  \nonumber  \\[1ex]    
&&+ \; {\bf H}^2 \; \tr \sum_{i_1=0}^2\cdots \sum_{i_{n}=0}^2
 \left(-\frac{1}{2}\right)^{ {N}_v(\vAi)} 
\left(-\frac{1}{2}\right)^{ \bar{N}_v(\vAbi)}  \nonumber  \\[1ex] 
&& \qquad \times T_{el}^\dagger(\vAbi)  cc T_{el}(\vAi) \nonumber  \\[1ex] 
&& + \; 2 \;{\cal R}e \; {\bf H} \; e^{-i \omega_0 z_L} \;   
\tr \sum_{i_1=0}^2 \cdots 
\sum_{i_{n}=0}^2   \left(-\frac{1}{2}\right)^{ \bar{N}_v(\vAbi)} 
\nonumber \\[1ex] 
&&\qquad \times T_{el}^\dagger(\vAbi) c \; \vAi(x,{\bf k};c)_{BG} \;.
\eeqar{contIS}

The first term in Eq.~(\ref{contIS}) is exactly the same as
for the case of asymptotic jets. For this, we directly 
use our results from the previous subsection. The second term in 
Eq.~(\ref{contIS})
cancels exactly at any order in opacity $n \geq 1$. The proof is 
again related to the multinomial decomposition of zero:
\beqar
&& {\bf H}^2 \; \tr \sum_{i_1=0}^2\cdots \sum_{i_{n}=0}^2
 \left(-\frac{1}{2}\right)^{ {N}_v(\vAi)} 
\left(-\frac{1}{2}\right)^{ \bar{N}_v(\vAbi)}  \nonumber  \\[1ex] 
&& \qquad \times T_{el}^\dagger(\vAbi)  cc T_{el}(\vAi) \nonumber  \\[1ex] 
&& = C_R \times C_R^n \;  {\bf H}^2  \left( -\frac{1}{2} 
-\frac{1}{2} +1  \right)^n    \equiv 0 \;.
\eeqar{zerp-pr}
The last term in Eq.~(\ref{contIS}) is the one that differentiates 
the case of initial-state medium-induced radiation from the 
 case of fully on-shell jets. We represent this term as follows:
\beq
\; 2 \; {\bf H} \cdot  \left( {\cal R}e  \; e^{-i \omega_0 z_L} 
\; {\bf J}_n  \;   \right) \;,
\eeq{nonhom}
where, similar to the case of ${\bf I}_{n-1}$ from Eq.~(\ref{bss2}), 
the  expression for  ${\bf J}_n$ can be simplified as follows: 
\beqar
&& {\bf J}_{n} =  \sum_{i_1\cdots i_{n}}
\left(-\half\right)^{N_v(\bar{\cal A}^{i_1\cdots i_{n}})}
 T_{el}^\dagger(\bar{\cal A}^{i_1\cdots i_{n}}) 
\nonumber \\[1ex]
&& \hspace*{1.5cm}\times  c \, {\cal A}_{i_1\cdots i_{n}} (x,{\bf k};c)_{BG}  
\nonumber \\[.5ex]
&& =  \sum_{i_1\cdots i_{n-1}} 
\left(-\half\right)^{N_v(\bar{\cal A}_{i_1\cdots i_{n-1}})}
 T_{el}^\dagger(\bar{\cal A}_{i_1\cdots i_{n-1}}) 
\nonumber \\[.5ex]
&&  \times \, \left[ C_A(e^{i(\omega_0-\omega_{n})z_{n}}
 e^{i{\bf q}_{n}\cdot\hat{\bf b}} -1) c
+[a_{n},c]\htB_{n} \right] \nonumber \\[1ex]
&& \times\, {\cal A}_{i_1\cdots i_{n-1}}  (x,{\bf k};c)_{BG}  \; .
\eeqar{bs2}
As in the case of Bertsch-Gunion radiation, the terms arising
from $\htB_n$ cancel for $ n > 1 $, see Eqs.~(\ref{bb}) and (\ref{bss2}).
The initial conditions for the iterative solution are the 
Bertsch-Gunion terms that come from the direct and virtual contributions:
\beqar   
{\bf J}_1  &=&  -    {\bf B}_{1} e^{i \omega_0 z_1} \; a_1 c [c, a_1] 
 - \frac{C_A}{2}  {\bf B}_{1} e^{i \omega_0 z_1} \; cc \nonumber \\[1ex]
  &=& - C_R C_A  {\bf B}_{1} e^{i \omega_0 z_1} \;.
\eeqar{incondJ}
The solution for  ${\bf J}_{n}$ can then be expressed as follows:
\beqar   
{\bf J}_n  &=&  
- C_R C_A^n \prod_{j=2}^n \left[ (e^{i(\omega_0-\omega_{j})z_{j}}
 e^{i{\bf q}_{j}\cdot\hat{\bf b}} -1) \right] \; \nonumber \\[1ex]
&& \times\;{\bf B}_{1} e^{i \omega_0 z_1} \; .
\eeqar{solJ}
Substituting Eq.~(\ref{solJ}) in  Eq.~(\ref{nonhom}) and recalling the
integration over the distribution of the momentum transfers from the
medium, we simplify the new non-homogeneous term in the solution
for the medium-induced bremsstrahlung:
\beqar
&& \left[ \prod_{i=1}^n  \int d^2 {\bf q}_i \; 
\frac{1}{\sigma_{el}} \frac{d\sigma_{el}}{d^2 {\bf q}_i } \right]  
\Bigg[ -  2  \, C_R C_A^n  \;  {\bf H} \cdot \Bigg( {\cal R}e
e^{-i \omega_0 z_L}    \nonumber  \\[1ex]
&&  \times  \prod_{j=2}^n 
\left[ (e^{i(\omega_0-\omega_{j})z_{j}}
 e^{i{\bf q}_{j}\cdot\hat{\bf b}} -1) \right] \;{\bf B}_{1} 
e^{i \omega_0 z_1} \Bigg)\; \Bigg]  \nonumber \\[1ex]
&& =  \left[ \prod_{i=1}^n  \int d^2 {\bf q}_i \; 
\left( \frac{1}{\sigma_{el}} 
\frac{d\sigma_{el}}{d^2 {\bf q}_i } -1 \right) \right] 
\Bigg[ -  2  \, C_R C_A^n  \;  {\bf H} 
\nonumber  \\[1ex]
&&  \cdot \left( {\cal R}e \, e^{-i \omega_0 z_L}
\prod_{j=2}^n  e^{i(\omega_0-\omega_{j})z_{j}}
e^{i{\bf q}_{j}\cdot\hat{\bf b}}  \;{\bf B}_{1} 
e^{i \omega_0 z_1} \right)\; \Bigg]   \nonumber  \\[1ex]
&& =  \left[ \prod_{i=1}^n  \int d^2 {\bf q}_i \; 
\left( \frac{1}{\sigma_{el}} 
\frac{d\sigma_{el}}{d^2 {\bf q}_i } -1 \right) \right] 
\bigg[ -  2  \, C_R C_A^n  \;   \nonumber  \\[1ex]
&&  \times {\bf H} \cdot {\bf B}_{(2 \cdots n)(1\cdots n)} 
{\cal R}e  \, e^{-i \sum_{j=2}^{n+1} \omega_{j \cdots n} \Delta z_j }
\bigg] \; .
\eeqar{sumhonh}
Here, we have denoted $z_{n+1} = z_L$ and 
$\omega_{n+1,n} \equiv \omega_0$.  With this solution for the
inhomogeneous term, we can now write the solution for initial 
state energy loss as the sum of the Bertsch-Gunion case and the 
destructive interference term, note the ``$-$'' sign
in Eq.~(\ref{sumhonh}). The net result reads: 
\begin{widetext} 
\beqar
k^+ \frac{dN^g(IS)}{dk^+ d^2 {\bf k} } &=& 
\frac{C_R \alpha_s}{\pi^2}
\left[ \prod_{i = 1}^n \int
\frac{d \Delta z_i}{\lambda_g(z_i)}  \right] 
\left[ \prod_{j=1}^n \int d^2 {\bf q}_j 
\left( \frac{1}{\sigma_{el}(z_j)} 
\frac{d \sigma_{el}(z_j) }{d^2 {\bf q}_j}   
-  \delta^2 ({\bf q}_j) \right)    \right]  \nonumber \\  
&& \left[  \; | {\bf B}_{(2\cdots n)(1\cdots n)} |^2    
+  2\, {\bf B}_{(2\cdots n)(1\cdots n)} \cdot 
  \sum_{i=2}^n {\bf B}_{(i+1 \cdots n)(i\cdots n)} 
\cos \left( \sum_{j=2}^i \omega_{(j\cdots n)} \Delta z_j \right) 
\right.  \nonumber \\  && \left.  
-  2\,  {\bf H} \cdot {\bf B}_{(2\cdots n)(1\cdots n)} 
\cos \left( \sum_{j=2}^{n+1} \omega_{(j\cdots n)} \Delta z_j \right) 
 \right] \;. \qquad 
\eeqar{ISfull}
To second order in opacity, direct comparison is made 
in Appendix~\ref{is2ord} with 
the brute force calculation. 
For the case of initial-state energy loss, $z_0$ is the position 
where the jet enters the medium and $z_L$ is the position of the 
hard interaction.
Writing  the color current propagators and 
interference phases directly, we obtain:
\beqar
k^+ \frac{dN^g(IS)}{dk^+ d^2 {\bf k} } &=& \frac{C_R \alpha_s}{\pi^2}
\left[ \prod_{i = 1}^n \int
{d \Delta z_i}\; { \sigma_{el}(z_i) \rho(z_i)}  \right] 
\left[ \prod_{j=1}^n \int d^2 {\bf q}_j    
 \left( \frac{1}{\sigma_{el}(z_j)} 
\frac{d \sigma_{el}(z_j) }{d^2 {\bf q}_j}   
-  \delta^2 ({\bf q}_j) \right)    \right]  \nonumber \\ 
&& \hspace*{-0.7cm} \times \left[ 
\left( \frac{{\bf k} - {\bf q}_2 - \cdots - {\bf q}_n}
{({\bf k} - {\bf q}_2 - \cdots - {\bf q}_n)^2}  - 
 \frac{{\bf k} - {\bf q}_1 - \cdots - {\bf q}_n}
{({\bf k} - {\bf q}_1 - \cdots - {\bf q}_n)^2}     \right)^2 
+ 2 \, \left( \frac{{\bf k} - {\bf q}_2 - \cdots - {\bf q}_n}
{({\bf k} - {\bf q}_2 - \cdots - {\bf q}_n)^2}  - 
 \frac{{\bf k} - {\bf q}_1 - \cdots - {\bf q}_n}
{({\bf k} - {\bf q}_1 - \cdots - {\bf q}_n)^2}     \right) \right.
 \nonumber \\ 
&& \hspace*{-0.7cm}  \cdot   \sum_{i=2}^n 
\left( \frac{{\bf k} - {\bf q}_{i+1} - \cdots - {\bf q}_n}
{({\bf k} - {\bf q}_{i+1} - \cdots - {\bf q}_n)^2}  - 
 \frac{{\bf k} - {\bf q}_i - \cdots - {\bf q}_n}
{({\bf k} - {\bf q}_i - \cdots - {\bf q}_n)^2}     \right) 
\cos \left( \sum_{j=2}^i  
\frac{ ({\bf k} - {\bf q}_j - \cdots - {\bf q}_n)^2 }{k^+} \Delta z_j   
 \right)    \nonumber \\ 
&&  \hspace*{-0.7cm} \left. - 2 \; \frac{{\bf k}}{{\bf k}^2}  \cdot 
\left( \frac{{\bf k} - {\bf q}_2 - \cdots - {\bf q}_n}
{({\bf k} - {\bf q}_2 - \cdots - {\bf q}_n)^2}  - 
 \frac{{\bf k} - {\bf q}_1 - \cdots - {\bf q}_n}
{({\bf k} - {\bf q}_1 - \cdots - {\bf q}_n)^2}     \right)
\cos \left( \sum_{j=2}^{n+1}  
\frac{ ({\bf k} - {\bf q}_j - \cdots - {\bf q}_n)^2 }{k^+} \Delta z_j  
 \right)    \right] \;.
\eeqar{ISfull-det}
\end{widetext}

\section{Numerical results}

In this section we carry out numerical simulations of the 
different radiative energy loss regimes, Eqs.~(\ref{full-final}), 
(\ref{BGfull}) and (\ref{ISfull}), to first order in 
opacity. The momentum transfers from the medium are given by the
$2 \rightarrow 2$  differential scattering cross section 
calculated in the Born approximation using a finite range, 
$\sim 1/\mu$,  Yukawa potential:
\beq  
\frac{d \sigma_{el}}{d^2 {\bf q}} = 
\frac{C_R C_T}{d_A} \frac{4 \alpha_s^2}{\pi}
\frac{1}{ ({\bf q}^2 + \mu^2)^2 } \;. 
\eeq{born}
Here, $C_R$ and $C_T$ are the quadratic Casimirs of the
jet and target representations and $d_A$ is the 
dimension of the adjoint representation. For QCD,
\beq
\frac{C_R C_T}{d_A} = \frac{9}{8}, \; \frac{1}{2}\;, \frac{2}{9}  \; ,
\eeq{casims} 
for $gg\rightarrow gg$, $q(\bar{q})g\rightarrow q(\bar{q})g$ and
$q(\bar{q}) q(\bar{q}) \rightarrow q(\bar{q}) q(\bar{q})$, 
respectively.

The total scattering cross section $\sigma_{el}$ can be absorbed 
in the mean free path, $\lambda_g = 1 /  \sigma_{el} \rho$, 
and the  normalized momentum transfer distribution from the 
medium is given by: 
\beq 
\frac{1}{\sigma_{el}}  \frac{d\sigma_{el}}{d^2 {\bf q}} = 
\frac{\mu_{\rm eff}^2}{\pi ({\bf q}^2+\mu^2)^2}\;, \quad
\mu_{\rm eff}^2 = \mu^2 \times \frac{\mu^2 + Q_{\max}^2}{Q_{\max}^2} \;.   
\eeq{qdist}
In Eq.~(\ref{qdist}), $\mu_{\rm eff}^2$ arises from the finite
range of integration $ {\bf q}^2 \leq Q_{\max}^2 =  s/4 = 
\mu E_{\rm jet} / 2 $.  This constraint is relevant  for small 
jet energies.

Next, use the results to first order in opacity derived here.
The reason for this $n=1$  choice is twofold: firstly, the 
formulas are simple enough to allow analytic insight in the
different behavior of initial- and final-state $\Delta E_{\rm rad}$. 
Secondly, the $n = 1$ Bertsch-Gunion result is the prototypical 
medium-induced radiative energy loss in QCD which sets the scale
relative to which the destructive LPM interference effects have to be
evaluated. For final-state energy loss it has also been demonstrated 
numerically that the the sum of all contributions up 
to $3^{\rm rd}$ order in opacity is well approximated 
by the $n=1$ term~\cite{Gyulassy:2000fs}.
From Eqs.~(\ref{full-final}), (\ref{finalfull-det}),    
(\ref{BGfull}), (\ref{BGfull-det}),    
 (\ref{ISfull}) and  (\ref{ISfull-det}) in a medium 
of constant density we have:
\begin{widetext}
\begin{eqnarray}
k^+ \frac{dN^g(BG)}{dk^+ d^2 {\bf k}} &=&  \frac{C_R \alpha_s}{\pi^2} 
 \int_0^L  \frac{d \Delta z}{\lambda_g}  
\int_{{\bf q}^2 \leq s/4}  d^2 {\bf q} \;
 \frac{1}{\sigma_{el}}  
\frac{d\sigma_{el}}{d^2 {\bf q}} \; \left[ |{\bf B}_1|^2  \right]  
\nonumber \\
&=& \frac{C_R \alpha_s}{\pi^2}  \frac{L}{\lambda_g} 
 \int_{{\bf q}^2 \leq s/4}   d^2 {\bf q} \;  
\frac{\mu_{\rm eff}^2}{\pi ({\bf q}^2+\mu^2)^2} 
\; \left[  \frac{{\bf q}^2}{{\bf k}^2 ({\bf k}-{\bf q})^2}   
  \right]  \;, \label{BG1} \\[2ex]
k^+ \frac{dN^g(IS)}{dk^+ d^2 {\bf k}} &=&  \frac{C_R \alpha_s}{\pi^2} 
 \int_0^L  \frac{d \Delta z}{\lambda_g} 
\int_{{\bf q}^2 \leq s/4}  d^2 {\bf q} \;
 \frac{1}{\sigma_{el}}  
\frac{d\sigma_{el}}{d^2 {\bf q}} \;  
\left[  |{\bf B}_1|^2 - 2 {\bf H} \cdot  {\bf B}_1 
\cos (\omega_0 (L-\Delta z) ) \right] 
\nonumber \\
&=& \frac{C_R \alpha_s}{\pi^2}  
\int_{{\bf q}^2 \leq s/4}  d^2 {\bf q} \;  
\frac{\mu_{\rm eff}^2}{\pi ({\bf q}^2+\mu^2)^2} 
\left[   \frac{L}{\lambda_g}   \frac{{\bf q}^2}{{\bf k}^2 
({\bf k}-{\bf q})^2}   
-  2    \frac{{\bf q}^2 - 
{\bf k} \cdot {\bf q} }{{\bf k}^2 ({\bf k}-{\bf q})^2}    
  \frac{k^+}{ {\bf k}^2 \lambda_g}  
\sin \left( \frac{  {\bf k}^2  L}{k^+} \right)   \right]  \;, \label{HBG1}
\\[2ex]
k^+ \frac{dN^g(FS)}{dk^+ d^2 {\bf k}} &=&  \frac{C_R \alpha_s}{\pi^2} 
 \int_0^L  \frac{d \Delta z}{\lambda_g} 
\int_{{\bf q}^2 \leq s/4}  d^2 {\bf q} \;  \frac{1}{\sigma_{el}}  
\frac{d\sigma_{el}}{d^2 {\bf q}} \;  \left[  - 2 {\bf C}_1 \cdot  {\bf B}_1 
\left( 1 - \cos\left( \omega_1 \Delta z \right) \right)  \right] 
 \nonumber \\
&=& \frac{C_R \alpha_s}{\pi^2}  
  \int_{{\bf q}^2 \leq s/4}  d^2 {\bf q} \; 
 \frac{\mu_{\rm eff}^2}{\pi ({\bf q}^2+\mu^2)^2} 
\left[  \frac{ 2 {\bf k} \cdot {\bf q} }{{\bf k}^2 ({\bf k}-{\bf q})^2}   
\left(   \frac{L}{\lambda_g} -
  \frac{k^+}{ ({\bf k}-{\bf q})^2\lambda_g}  
\sin \left( \frac{ ({\bf k}-{\bf q})^2 L}{k^+}  \right) 
 \right)   \right]  \;, 
\label{CBG1}
\end{eqnarray}

\end{widetext}

Let us examine qualitatively the behavior of the radiative
spectrum and the total lightcone momentum loss, $\Delta p^+$. 
For energetic jets ($E_{\rm jet}/ m_q \rightarrow \infty$) we 
use ``energy loss'' and ``lightcone
momentum loss'' interchangeably. The magnitude of
$\Delta p^+$ depends on the available phase space for the 
radiative gluon at large  $k^+$. For the case of Bertsch-Gunion 
radiation, Eq.~(\ref{BG1}), $k^+$ does not appear in the integrand.
Therefore, in addition to the linear dependence on the path 
length $L$, the energy loss is proportional to the available 
energy. In Ref.~\cite{Gunion:1981qs} this can be seen through the 
flat rapidity dependence of ${dN^g(BG)}/{dy d^2 {\bf k}}$,
since $ y = \ln k^+/p^+$. 
The result is qualitatively similar to bremsstrahlung energy 
loss in QED, see Eq.~(\ref{radQED}), and we find:
\begin{equation}       
\Delta p^+(BG) \sim C_R \alpha_s p^+  \frac{L}{\lambda_g} 
\times f_{BG}(\mu, Q_0, p^+) \;. 
\label{BGqual}
\end{equation}
In Eq.~(\ref{BGqual}) $f_{BG}(\mu, Q_0, p^+)$ depends 
on the momentum transfers from the medium, the mass scale 
in the monopole nucleon
form-factor, $Q_0$, and may have additional weak logarithmic
dependence on $p^+$ via $Q_{\max}^2$.

The qualitative behavior of initial-state energy loss can be 
understood in the limit of large $p^+$ when 
$ {\bf k}^2 L / k^+ \ll 1$.
Expanding the $\sin(\cdots)$ term in Eq.~(\ref{HBG1}), we find 
\begin{equation}       
\Delta p^+(IS) \sim C_R \alpha_s p^+  \frac{L}{\lambda_g} 
\times f_{IS}(\mu, Q_0, p^+) \;, 
\label{ISqual}
\end{equation} 
with $f_{IS}(\mu, Q_0, p^+) \ll f_{BG}(\mu, Q_0, p^+)$.  
The difference in the overall magnitude of $\Delta p^+$ arises from 
the cancellation of part of the Bertsch-Gunion radiation through its 
interference with the bremsstrahlung from the hard collision. Still,
in the asymptotic regime $\Delta E_{\rm rad}$ remains proportional to 
the jet energy and depends linearly on the size of the nuclear matter. 
This result is qualitatively and quantitatively different from 
the argument given in Ref.~\cite{Baier:2000mf} for the Bertsch-Gunion
case in the limit of a very large number of soft scatterings, namely that 
initial-state $\Delta E_{\rm rad}$ is independent of the energy, grows 
quadratically with $L$ and is smaller than final-state energy loss  
by a factor of three. The reason for this difference is twofold: 
firstly, as indicated above initial-state energy loss was approximated 
by the case of on-shell jets $ \pm \infty $ in~\cite{Baier:2000mf}. 
Secondly, while our approach is general enough to address both 
$L/\lambda_g \sim {\rm few} $ and  $L/\lambda_g \gg {\rm few} $ 
cases, the numerical results presented here address the limit of 
few scatterings which we anticipate is relevant for finite nuclei. 
Whether the distinction between initial- and final-state energy loss
will be reduced for very large number of jet-medium interactions 
is still an open question.

Final-state non-Abelian energy loss has been analyzed in detail 
in~\cite{Gyulassy:2000er,Gyulassy:2000fs}. {\em Naively}, one may expect
that in the high energy regime the expansion of the $\sin(\cdots )$ 
term in the integrand of Eq.~(\ref{CBG1}) may lead to 
an $\sim L^3$ dependence of the radiative $\Delta E_{\rm rad}$. However, 
when properly weighted by the available phase space for 
the emitted gluon, this is reduced to a quadratic dependence, 
at most. In addition,
the extra powers of $1/k^+$ cancel the linear dependence
on the of the energy loss on $E_{\rm jet}$, leaving only a logarithmic
dependence. Thus,
\begin{equation}   
\Delta p^+(FS) \sim C_R \alpha_s   \frac{ \mu^2 L^2 }{\lambda_g} 
\times f_{FS}(\mu, Q_0, p^+) \;. 
\label{FSqual}
\end{equation}

\begin{figure}[!b]
\begin{center}
\vspace*{.5in}
\epsfig{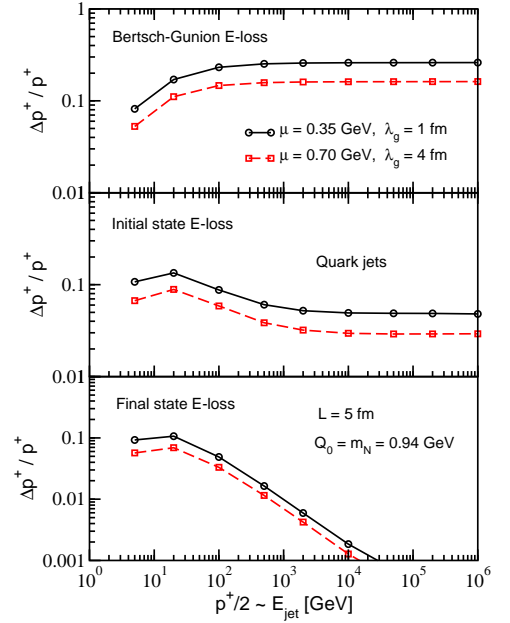}
\vspace*{.1in}
\caption{(Color online) Fractional energy loss for massless quark 
jets versus
the jet energy, $E_{\rm jet}$, in cold nuclear matter of length 
$L= 5$~fm.  Two different sets of typical momentum transfer
per scattering and gluon mean free path have been used for
comparison: $(\mu, \lambda_g) = (0.35$~GeV$, 1$~fm$)$ and  
$(\mu, \lambda_g) = 
(0.7$~GeV$, 4$~fm$)$. }
\label{fig-DEovE}
\end{center} 
\end{figure}

To study the radiative energy loss quantitatively, we have to identify the 
$|{\bf k}|$ scale, at which non-perturbative effects can become
important. For cold nuclear matter, in the original work of Bertsch and 
Gunion~\cite{Gunion:1981qs},  this scale was approximated by 
$Q_0 = m_\rho \simeq 770$~MeV. 
In generalized vector dominance models, the effective mass 
is bigger, $Q_0 \simeq 1$~GeV, due to the contribution of the 
heavier vector meson states. One can also understand 
the appearance of form-factors in a purely partonic picture. 
As $ |{\bf k}| \rightarrow 0 $, the transverse 
size of the radiative gluon $A_\perp =  1 / {\bf k}^2 $ exceeds 
the target parton  size and the strength of the interaction of the 
jet + gluon system is limited. For cold nuclear matter $1/Q_0 \leq 1/\mu 
\leq 1 / \Lambda_{QCD}$. We take $Q_0 = m_N = 0.94$~GeV from 
the independent calculation of dynamical nuclear 
shadowing~\cite{Vitev:2006bi,Qiu:2004da}, suggestive of 
partonic spots of size $\sim 0.2$~fm inside the nucleon. 
We implement this mass scale as follows:
\begin{equation}   
{\bf k}^2 \rightarrow {\bf k}^2  + Q_0^2 \;\;  (+ x^2 M_q^2) \;.
\label{ff}
\end{equation}    
The last term in Eq.~(\ref{ff}) arises for heavy quarks 
and was discussed in~\cite{Djordjevic:2003zk}. 
In addition, at any fixed order the opacity expansion series
is not a perfect square of a sum of amplitudes. Still, the
integrands in Eqs.~(\ref{HBG1}) and (\ref{CBG1}) can be 
represented as  $( |{\bf B}_1|^2 - {\bf correction} ) $. 
The requirement that the interference does not cancel differentially 
more than the available bremsstrahlung, induced by the medium 
in the absence of coherence, can be formulated as follows
to any order in opacity:
\begin{equation}
\sum_{i = 1}^{n} k^+ \frac{dN^g_{(i)}}{dk^+ d^2 {\bf k}}  \geq 0 \;.
\label{pos}
\end{equation}
Eq.~(\ref{pos}) can also used to identify and eliminate the parts 
of phase space where the approximations that we made are least 
reliable (e.g., the part of phase space where a gluon emitted 
in the direction opposite to the momentum transfer from the medium).
To set the kinematic limits, we require that the positive gluon energy
$ k^+ \geq Q_0$ and the gluon rapidity,
\begin{equation}
 y_g = \frac{1}{2} \ln \frac{k^+}{k^-} = 
 \frac{1}{2}\ln \frac{(k^+)^2}{{\bf k}^2}  \;,   
\label{yg}
\end{equation}
be within the rapidity gap from the target to the projectile,
$0 \leq y_g  \leq  \ln \, (p^+_{\rm proj/tar} / m_N ) $. 
Such kinematic constraints correspond to 
\begin{eqnarray}
\frac{Q_0}{p^+}  \leq x = \frac{k^+}{p^+} \leq 1 
\label{xcons}\;, \\
{x \, m_N}   \leq  |{\bf k}|  \leq  k^+ \; .
\label{kcons}
\end{eqnarray}
We have checked that in the limit of large $p^+$ there is 
little or no sensitivity of the energy loss to 
a factor of two increase or decrease  of the lower bounds in 
Eqs.~(\ref{xcons}) and (\ref{kcons}).

In Fig.~\ref{fig-DEovE} we show numerical results for 
the fractional lightcone momentum loss $\Delta p^+/p^+$ 
of massless quark jets in cold nuclear matter of length $L = 5$~fm.  
For the reference Bertsch-Gunion case, Eq.~(\ref{BG1}),  
in the limit of large jet energies
we recover $\Delta p^+ \propto p^+$. We study two sets 
of momentum transfers and mean free paths describing  
cold nuclear matter: $(\mu_1 = 0.35\; {\rm GeV}, \; 
\lambda_{g, \,1} = 1\;{\rm fm })$ and  $(\mu_2 = 0.7\; 
{\rm GeV}, \; \lambda_{g, \,2} = 4\;{\rm fm })$. Clearly,
the energy losses for these cases differ by close to a 
factor  of two,  indicating that  
$(\mu^2/ \lambda_{g })_1  = ( \mu^2/ \lambda_{g })_2$    
is not a universal parameter responsible for the quenching
of jets.  It should be noted that  for only a few 
scatterings in the medium with small momentum transfer 
$\propto \mu^2 = 0.12 \; {\rm GeV}^2$ quarks can lose 
on the order of 30\% of their energy.

The physically relevant case of initial-state energy loss,
Eq.~(\ref{HBG1}), is shown in the middle panel of Fig.~\ref{fig-DEovE}.  
Destructive interference   effects can lead to a large  
six-fold reduction of the fractional energy loss 
$\Delta p^+/p^+$  for  $p^+ \rightarrow \infty$. Still, qualitatively 
the behavior is similar to that of the Bertsch-Gunion case discussed
above. For medium parameters  $(\mu_1 = 0.35\; {\rm GeV}^2\;, \; 
\lambda_{g, \,1} = 1\;{\rm fm })$, directly relevant to  
phenomenology~\cite{prep}, quark jets lose $\sim 5\%$
of their energy. In contrast, for final-state energy loss,
shown in the bottom panel of Fig.~\ref{fig-DEovE}, the 
cancellation of the radiation in the asymptotic domain 
is still effective, leading to $\Delta p^+/p^+ \rightarrow 0$.
Although there is a  30\%  difference between the two 
calculations, only in this regime of final-state radiation
in cold nuclear matter may $\mu^2/\lambda_g$ be considered 
an {\em approximately} relevant jet quenching parameter.

\begin{figure}[!tb]
\begin{center}
\epsfig{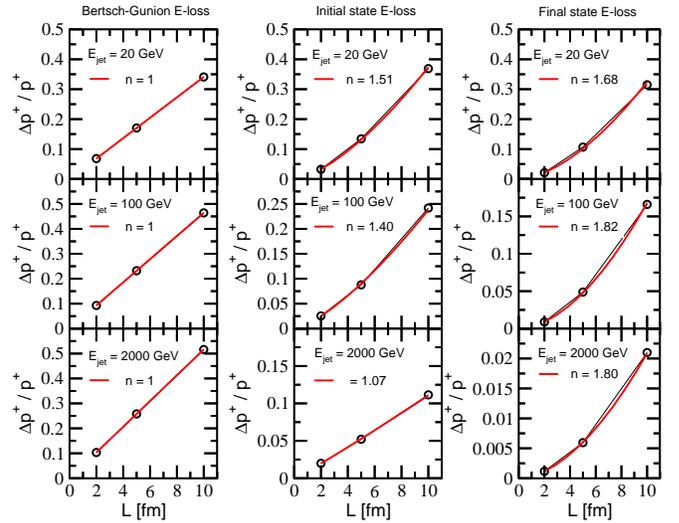}
\vspace*{.1in}
\caption{(Color online) Fractional energy loss for massless 
quark jets of 
$ E_{\rm jet} = 20$~GeV,  $100$~GeV and $2000$~GeV. Three different
path lengths $L = 2$~fm, $5$~fm and $10$~fm were used.  
Power law fits indicate the system size dependence of 
$\Delta p^+ / p^+  = \Delta E_{\rm rad} / E$.  } 
\label{fig-EEovElen}
\end{center} 
\end{figure}

Numerical results for the path length dependence of jet 
energy loss in the three different regimes, Eqs.~(\ref{BG1}), 
(\ref{HBG1}) and  (\ref{CBG1}),  are presented 
in Fig.~\ref{fig-EEovElen} for three jet 
energies, $E_{\rm jet} = 20,\; 100, \;2000$~GeV, respectively. 
By definition, the reference Bertsch-Gunion case without 
interference yields a linear 
dependence of $\Delta p^+/ p^+$ on the path length $L$. 
Numerical calculations were carried out for three different 
path lengths, $L= 2, \; 5 , \; 10$~fm, and power law fits 
were used to extract the power index $n$ of the $\Delta p^+ 
\propto L^n$ dependence.  At large $p^+$ initial-state 
energy loss approaches  a  linear behavior, $n=1$, 
while final state energy loss shows approximately 
quadratic behavior, $n = 2$, for static nuclear 
matter. 

%%%%%****************************

Both results have important implications for heavy 
ion phenomenology. Let us define the nuclear 
modification factor, 
\begin{eqnarray} 
R_{AB}(p_T) & = &  \frac{ d N^h_{AB}/dyd^2 p_T }
{ T_{AB} \; d \sigma^h_{pp}  / dyd^2 p_T } \; ,  
\label{raa}
\end{eqnarray} 
which is used to identify dense matter effects on
parton propagation. In Eq.~(\ref{raa}), $T_{AB}$ 
is the nuclear overlap function, which relates 
the hadron multiplicity in the many-body collision
to the corresponding cross section in p+p reactions.
In A+A collisions, the quadratic dependence
of final-state energy loss is reduced to linear, 
when Bjorken expansion is taken into account, and 
we predict the following suppression pattern:
\begin{equation} 
\ln R_{AA}(p_T) = -\kappa_{AA} N_{\rm part}^{2/3} \;.
\label{RAAqgp}
\end{equation} 
In Eq.~(\ref{RAAqgp})  $N_{\rm part}$ is the total number of
participants and $\kappa_{AB}$ is a  microscopic
coefficient, which depends on the properties of the dense
latter.
Initial-state energy loss, on the other hand, approaches
linear dependence on the system size 
$\sim N_{\rm part}({\rm target})$, which can be used
to predict analytically~\cite{prep} the centrality 
dependence of forward  rapidity hadron suppression in p+A 
reactions~\cite{Vitev:2006bi}. 
Here $N_{\rm part}({\rm target})$ is the (average) 
number of nuclear target participants along the projectile 
line for a given (centrality class) impact parameter.  
The quantitatively different behavior of initial-state 
energy loss leads to the following result in p+A reactions:
\begin{equation}
\ln R_{pA} (p_T)  = 
- \kappa_{pA}  N_{\rm part}({\rm target}) \;.
\label{RAAcold}
\end{equation}

Before we proceed to more differential bremsstrahlung distributions
it is important to emphasize that the quadratic dependence
of final-state energy loss on the size of the static medium, 
$\propto L^2$, does not imply that the magnitude of the 
energy loss itself is large. On the contrary, such 
dependence arises from the maximally efficient cancellation 
of the large $x = k^+ / p^+$ radiation and  for the same
parameters of the medium, in the limit of $p^+ \rightarrow \infty$, 
final-state energy loss is negligible when compared to 
initial-state energy loss.

\begin{figure}[!tb]
\begin{center}
\epsfig{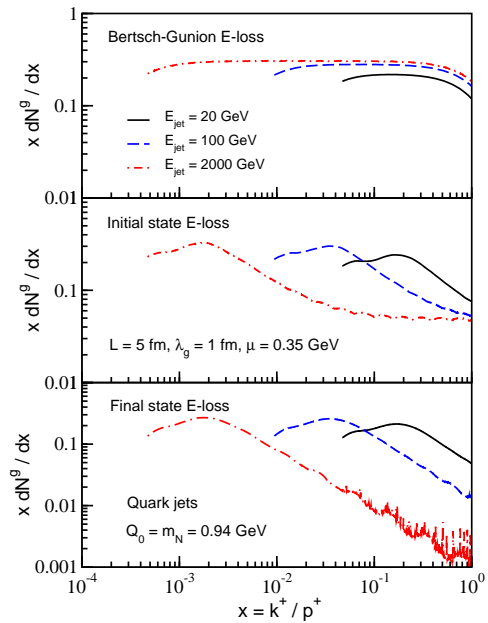}
\vspace*{.1in}
\caption{(Color online) Radiation intensity 
$x d N^g/dx$ for $E_{\rm jet} = 20$~GeV,  
$100$~GeV and  $2000$~GeV quark jets. Note the difference to first 
order in opacity in the large $x = k^+/p^+$  behavior for 
Bertsch-Gunion, initial-state and final-state bremsstrahlung. }
\label{fig-xdNdX}
\end{center} 
\end{figure}

One can gain insight in the energy dependence of 
$\Delta p^+$ by investigating the radiative intensity 
spectrum $xdN^g/dx$, obtained by integrating the double 
differential distributions Eqs.~(\ref{BG1}), (\ref{CBG1})   
and (\ref{HBG1}) over ${\bf k}$  
and  shown in  Fig~\ref{fig-xdNdX}. We 
have chosen $L=5$~fm  $ (\mu_1 = 0.35\; {\rm GeV}, \; 
\lambda_{g, \,1} = 1 \; {\rm fm }) $ and the same jet 
energies as in Fig.~\ref{fig-EEovElen}. For the 
Bertsch-Gunion case there is an approximately flat,
up to the edge of phase space constraints, dependence 
of the bremsstrahlung intensity on $x$, leading 
to $\Delta p^+/p^+ \approx {\rm const}_{BG}. $ 
For initial-state energy loss there is a finite cancellation
of the  intensity spectrum at large values of $k^+$, which
leads to a suppressed $ \Delta p^+/p^+  \approx {\rm const}_{IS}$. 
For final-state energy loss, the spectrum is progressively 
more suppressed by destructive interference effects 
that lead to a power behavior of the intensity spectrum, 
$xdN^g/dx \propto 1/k^+$. Therefore, in this
case, energy loss depends logarithmically  on the jet 
energy, as first derived 
in~\cite{Gyulassy:2000er,Gyulassy:2000fs}.

\begin{figure}[!tb]
\begin{center}
\epsfig{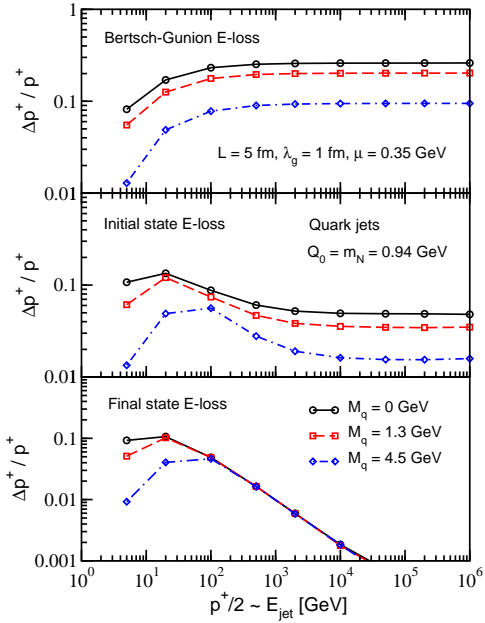}
\vspace*{.1in}
\caption{(Color online) Mass dependence of the fractional energy loss 
$\Delta E_{\rm rad} / E$  versus the jet energy, $E_{\rm jet}$. 
We studied
massless quarks, $M_q = 0$~GeV, charm quarks  $M_c = 1.3$~GeV
and bottom quarks  $M_b = 4.5$~GeV. }
\label{fig-DEovEmass.eps}
\end{center} 
\end{figure}

The intensity spectrum, shown in  Fig~\ref{fig-xdNdX},  is also 
important for understanding the effect of a heavy quark mass,
$M_q$, on the magnitude of the energy loss, 
$\Delta E_{\rm rad}$. Recall 
that the modulation of the 2D propagators and interference
phases is $\propto x^2 M_q^2$, see Eq.~(\ref{ff}) and 
Ref.~\cite{Djordjevic:2003zk}.  For the bremsstrahlung intensity,
which has a negligible contribution in the region of large $x$
where the mass correction is significant, we expect that 
there will be no difference between $\Delta E_{\rm rad}$  
for light and heavy quark. In contrast, if 
the kinematically allowed region in $x = k^+/p^+$ contributes
equally to the radiation intensity, as in the Bertsch-Gunion
case, or the cancellation of the large $x$ radiation is partial, 
as in the case of initial-state energy loss, then the dependence
of $\Delta E_{\rm rad}$ on the heavy quark mass should remain. 
Numerical results for the three cases discussed in this 
paper are shown in Fig.~\ref{fig-DEovEmass.eps}. We used
the same choice of parameters that describe cold nuclear 
matter in Fig.~(\ref{fig-xdNdX}). Quark 
masses for light, charm, and bottom quarks have been
set to: $M_{u,d} = 0$~GeV, $M_{c} = 1.3$~GeV, and 
$M_{b} = 4.5$~GeV.

It is seen that, for final-state energy loss in cold nuclear
matter, the charm quark $\Delta E_{\rm rad}$ becomes comparable 
to the light quark $\Delta E_{\rm rad} $ at  
$E_{\rm jet} \approx (p^+/2) = 20$~GeV. 
For the much heavier bottom quarks,  equality 
is reached at around 100~GeV. These results, given the crude
energy binning used here to cover a very large dynamic range,
$  5\; {\rm GeV} < p^+/2  <  10^6 \; {\rm GeV} $, are 
not inconsistent with similar findings for heavy quarks
traversing a quark-gluon plasma~\cite{Djordjevic:2003zk}.
On the other hand, for the Bertsch-Gunion case and the 
physically relevant case of initial-state energy loss,  
the mass dependence persists at any jet energy, see 
Fig.~\ref{fig-DEovEmass.eps}.

\begin{figure}[!t]
\begin{center}
\epsfig{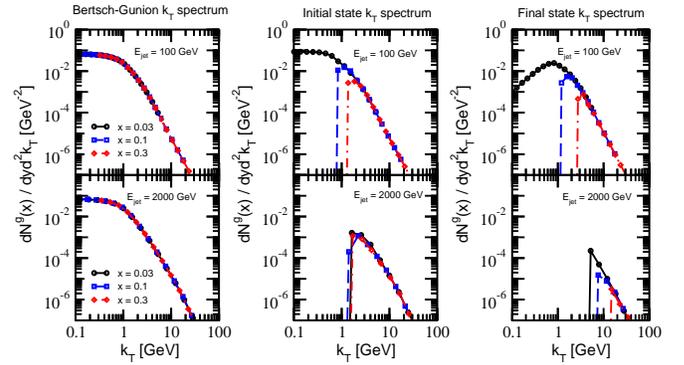}
\vspace*{.1in}
\caption{(Color online) Differential in the gluon transverse 
momentum $|{\bf k}|$ bremsstrahlung 
multiplicity distribution $dN^g/dy d^2 {\bf k}$   for 
lightcone momentum fractions $x = k^+/p^+  = 0.03, 0.1$ 
and $0.3$. We show for comparison quark jets of $E_{\rm jet} 
= 100$~GeV and $2000$~GeV 
and the Bertsch-Gunion, initial-state and final-state spectra for 
comparison. }
\label{fig-dNdYd2K}
\end{center} 
\end{figure}

The last numerical result, presented in this manuscript, 
is the fully differential gluon bremsstrahlung distribution
$dN^g/dyd^2{\bf k}$ for select values of $x = k^+/p^+$ versus 
the gluon transverse momentum. We have chosen  $x = 0.03, 0.1$ 
and $0.3$ to cover both small and large values of the gluon 
lightcone momentum fraction. Two different jet energies,
$E_{jet} = 100, 2000$~GeV are also shown in Fig.~\ref{fig-dNdYd2K}.     
Massless quarks were used as an example in this study. 
The most characteristic feature of the Bertsch-Gunion energy 
loss is that at large values of the momentum, $|{\bf k }|$,  
the bremsstrahlung spectrum behaves $ \sim 1/{\bf k}^4$.  Note 
that this behavior is different when compared to 
the $ \sim 1/{\bf k}^2$
for hard bremsstrahlung. The 
turnover in the growth of the spectrum at small $|{\bf k}|$   
is controlled by the largest of the non-perturbative scales
in the problem, in our case, $Q_0$. Note that there is no real
difference as a function of $x$ or $E_{\rm jet}$, as expected 
from Eq.~(\ref{BGfull}) to first order in opacity.

At the other extreme is the final-state medium-induced 
differential gluon distribution, see the right bottom panel 
of Fig.~\ref{fig-dNdYd2K}.  The cancellation 
of the small $|{\bf k}|$ radiation of the basic Bertsch-Gunion 
term clearly becomes more effective at larger values of $x$.  
Given the spectrum steeply falling in $|{\bf k}|$, this 
cancellation is reflected in the large $x$ suppression
seen in Fig.~\ref{fig-xdNdX} and the qualitatively different
behavior of the final-state $\Delta E_{\rm rad}$ when compared 
to QED or the Bertsch-Gunion limit. Initial-state energy loss  
exhibits a cancellation which, however, is limited at large 
values 
of $x$, see the middle bottom panel of Fig.~\ref{fig-dNdYd2K}. 
This explains the {\em finite} reduction of the energy loss 
relative to the incoherent limit of the medium-induced 
bremsstrahlung in QCD.

\section{Discussion and conclusions}

In this paper, in the framework of the reaction operator 
approach~\cite{Gyulassy:2000er}, we derived the full solution 
for the medium-induced gluon bremsstrahlung for 
asymptotic, $t = \pm \infty$, on-shell fast partons 
(Bertsch-Gunion) 
as well as  partons that undergo hard 
scattering  ($Q^2 \gg Q_0^2 \geq \mu^2 \geq \Lambda_{QCD}^2$ ) 
to produce high-$p_T$ or high-$E_T$  particles and jets 
(physically relevant case). The double differential intensity 
spectrum, $x dN^g/dx d^2{\bf k}$, was derived as an 
infinite opacity  series: an expansion in the 
correlation between the sequential  multiple scatterings 
in the medium. Our general results are  suitable for further 
analytic approximations and/or numerical simulations.

One of the main findings of this work is that, in contrast to
final-state energy loss, the cancellation of the 
Bertsch-Gunion radiation for initial-state $\Delta E _{\rm rad}$   
due to the non-Abelian LPM effect is less effective.  While there
can be a large (in our examples six-fold) reduction of the 
total $\Delta E _{\rm rad}$  relative to the incoherent 
Bertsch-Gunion case, such a cancellation does not 
qualitatively change  the path length, $\propto L$, and
the energy, $\propto E$, dependence of the  radiative energy loss
for finite nuclei in the case of few interactions. 
It is clearly the incoherent Bertsch-Gunion regime that sets
the upper limit for the amount of energy loss induced by strongly
interacting matter.  Of the two cases relevant to hard probe 
physics, it is initial-state energy loss that by far 
dominates over final-state energy loss for the same medium 
parameters, such as the gluon mean free path, $\lambda_g$, and
the squared momentum transfer, $\mu^2$.

The reason that final-state energy loss has been found 
to play a dominant role in A+A collisions is the large 
density of the QGP co-moving with the jet. Moreover, the 
effect of final-state $\Delta E_{\rm rad}$  is  amplified
by the steeply falling  spectra of  {\em outgoing} partons, 
$\sim 1/p_T^n$ with $n \geq 4$~\cite{Gyulassy:2003mc}. This 
observation  implies that, in a region where the flux 
of {\em incoming} 
partons is rapidly changing as  a function of the $p_T$/$E_T$ 
of the final-state particle/jet, the cold nuclear matter 
$\Delta E _{\rm rad}$ can lead to a significant and observable 
suppression of hadron production~\cite{Vitev:2006bi,Kopeliovich:2005ym}. Such 
regions of phase space in p+A and A+A reactions are the ones 
near the projectile rapidity. In such kinematic domains in the 
rest frame  of the  target nucleus the incoming partons are of 
almost asymptotically high energy. The knowledgeable reader 
can easily verify that for  $p_T = 1-2$~GeV  at rapidity  $y=4$, 
a region accessible by the STAR  collaboration 
at RHIC~\cite{Adams:2006uz},  the incoming partons 
are of energy $E \geq 20$~TeV. Therefore, the contribution 
to $\Delta E$  from final-state radiative energy 
loss and collisional energy loss is completely negligible.  
In fact, due to the rapidity boost, the incoming and outgoing 
parton energies are very large everywhere, except in the 
backward region near the target rapidity. Theoretical 
results, derived in this manuscript, can be used to develop
a more complete and consistent pQCD phenomenology of  
proton-nucleus collisions~\cite{prep}. Given the short 
time scale of dynamical cold nuclear matter effects, 
$ t_{\rm col}  \ll   t_{\rm eq}   \ll  L_T^{QGP}$, 
these must also be incorporated in the description of 
hard processes in A+A reactions.

We also argue that a new application of cold nuclear matter-induced  
radiative energy loss can be related to particle production at 
backward rapidities. It has been established, through the 
measurements of the PHENIX collaboration~\cite{Adler:2004eh} at
RHIC, that the forward rapidity hadron suppression in p+A 
reactions is correlated with a backward rapidity enhancement. 
There are no models at present that can consistently account 
for this effect. One possibility is that such enhancement 
in the inclusive particle yields may arise from the  induced  
gluon bremsstrahlung.  What makes such a scenario plausible is  
the fact that the gluon yield, $dN^g/dxd^2{\bf k}$, is dominated by 
small $x$ radiation and the destructive LPM effect cancels the 
small $|{\bf k}|$ part of the induced gluon spectrum.  For the
final-state $\Delta E_{\rm rad}$, this contribution is expected 
to yield an enhancement of the soft, large-angle hadrons associated 
with an away-side triggered jet~\cite{Vitev:2005yg}. For 
initial-state cold nuclear matter energy loss, this corresponds 
to gluons  emitted preferentially in the backward rapidity 
(or near target rapidity) region, see Eq.~(\ref{yg}). 
Since at moderate $k_T$ the power behavior 
of the gluon spectrum is similar to that of hard 
scattering, $\propto 1/k_T^4$,  it may have a sizable 
contribution to the inclusive particle yields.

We emphasize that the results derived in this paper are 
applicable to both cold nuclear matter and the quark-gluon plasma. 
However, the physics situation of asymptotic $- \infty$, $+ \infty$ 
jets propagating through the QGP  (Bertsch-Gunion)
or jets undergoing hard scattering at a finite time  $t = L$, 
having penetrated the QGP, (initial-state) cannot be experimentally 
realized at present. Nevertheless, the  formal radiative 
energy loss solutions  obtained here 
are general enough to describe both cases of  interest.

In summary, we anticipate that the results derived in this paper,
in particular the ones related to initial-state cold nuclear
matter radiative energy loss, will play an important role
in a consistent many-body pQCD description of hard processes 
in high energy reactions of heavy nuclei and complement 
the well-developed theory and phenomenology of final-state
energy loss in the quark-gluon plasma.

\begin{acknowledgments}
Useful discussion with T. Goldman is acknowledged. This work 
is supported by the U.S. Department of Energy under 
contract no. DE-AC52-06NA25396 and the J. Robert Oppenheimer 
Fellowship of the Los Alamos National Laboratory.
\end{acknowledgments}

\begin{appendix}
\section{ Simplifying the energy loss calculation for 
energetic partons}
\label{silmlify}

In this paper we adopt the notation: 
\beq
p^\mu = [p^+, p^-, {\bf p}] = \left[ p^0+p^-, p^0-p^- , {\bf p}  \right] \;,  
\eeq{lightcone} 
for the lightcone momentum components. The scalar product between
two 4-vectors is then given by:
\beq
p\cdot q = \frac{1}{2} \left( p^+ q^- + p^- q^+  \right) - 
{\bf p} \cdot {\bf q} \;.
\eeq{prod}
The soft gluon approximation that we use implies:
\beq
k^+ \ll p^+ \;
\eeq{plcomp}
for the positive lightcone momentum of the gluon relative 
to the lightcone momentum of the jet. Evidently, the gluon 
{\em multiplicity} should be dominated by small $x = k^+/p^+$
gluons, which we can verify {\em a posteriori} in all cases.
The gluon intensity spectrum, however, can have contributions
from the large $x$ region of phase space.
For the transverse momenta we require:
\beq
\frac{ ( {\bf k} - \sum _i {\bf q}_i )^2}{k^+} \ll k^+ \;. 
\eeq{trcomp}
This constraint implies that at the {\em emission vertex} the transverse 
momentum $ {\bf k} - \sum _i {\bf q}_i $ of the gluon is small 
and the large transverse momenta ${\bf k}$ are accumulated via
multiple scattering, which can be large angle. Finally, 
\beq
\frac{ ( {\bf k} - \sum _i {\bf q}_i )^2}{p^+} \approx 0 
\ll \frac{ ( {\bf k} - \sum _i {\bf q}_i )^2}{k^+}   \;, 
\eeq{trcomp1}
and such terms are neglected. 
The polarization vector for the physical final-state
gluon is given by 
\beq
\epsilon(k)=\left[ \, 0 , 2\frac{ \epsilon_\perp \cdot
{\bf k}}{k+}, \epsilon_\perp \, \right]\;\;, 
\eeq{kinemw}
so that $\epsilon(k) \cdot k = 0$.  With $n$ momentum transfers
from the medium and the approximations  outlined in Eqs.~(\ref{plcomp})
 - (\ref{trcomp1}),  the kinematic part of the gluon emission 
vertex reads:
\beq
\Gamma_{i_1\cdots i_n} \approx 2p^+ (k^+)^{n-1} 
\epsilon_\perp \cdot ({\bf k} - {\bf q}_{i_1} - \cdots - 
{\bf q}_{i_n} ) \; .
\eeq{kinvert}
The color factor, associated with Eq.~(\ref{kinvert}), 
is $(-f^{ca_nd_n})(-f^{d_na_{n-1}d_{n-1}})\cdots 
(-f^{d_2a_{1}d_{1}})ig_s \,t_{d_1}$.  Here, $t_{d_1} \equiv 
d_1$ is the color matrix at the emission vertex.
Thus, a common factor $- 2 i g_s \epsilon_\perp \cdot 
(\cdots) $ can be factored out from all amplitudes.

While the integrals over the longitudinal position of the 
scattering centers, $z_i$, have to be taken explicitly, 
a major simplification  at the level of squared amplitudes 
occurs when we consider the average over the corresponding  
impact  parameters 
${\bf b}_i={\bf x}_i-{\bf x}_0$. This
average is done in the {\em local rest frame} of the 
medium containing colored scattering centers. For a scattering 
center, $i$, that appears in a direct interaction the  impact 
parameter average takes the form: 
\begin{widetext}
\beqar
\langle \cdots \rangle_{A_\perp} &=& \langle \cdots
\int \frac{d^2{\bf b}_{i}}{A_\perp} \,
 (-i)\int\frac{d^2 {\bf q}_{i}}{(2\pi)^2} \,{\cal A}_{el}(0,{\bf q}_{i}) \,
 e^{-i{\bf q}_{i}\cdot{\bf b}_{i}} \, 
(+i)\int\frac{d^2 {\bf q}_{i}^{\prime}}{(2\pi)^2} 
\,{\cal A}_{el}^*(0,{\bf q}_{i}^{\prime}) \, e^{+i{\bf q}_{i}^{\; \prime}
\cdot{\bf b}_{i}}  \; \cdots \rangle \; \nonumber \\
\hspace*{-.5cm} &=&  \cdots \int \frac{d^2 {\bf q}_{i}}{(2\pi)^2} \,
\frac{d^2 {\bf q}_{i}^{\prime}}{(2\pi)^2} 
\frac{(2\pi)^2 \delta^2 ({\bf q}_{i} - 
{\bf q}_{i}^{\prime})}{A_\perp} 
{\cal A}_{el}^*({\bf q}_{i}^{\prime})   {\cal A}_{el}({\bf q}_{i}) 
 \; \cdots 
=  
\cdots \frac{\sigma_{el}}{A_\perp} 
\int \frac{d^2 {\bf q}_{i}}{(2\pi)^2} \left[ \frac{1}{\sigma_{el}}
\frac{d\sigma_{el}}{d^2{\bf q}_i}  \right] \,
\; \int d^2 {\bf q}_{i}^{\prime} 
\delta^2 ( {\bf q}_{i} -  {\bf q}_{i}^{\prime} ) 
\; \cdots \, \;\; , \nonumber \\
&&
\label{travd}
\eea
which  constrains the transverse momentum exchanges 
${\bf q}_{i} = {\bf q}_{i}^{\prime}$  to be equal in the 
amplitude and its conjugate. For a double-Born  interaction:  
\beqar
\langle \cdots \rangle_{A_\perp} &=&  \langle \cdots
\int \frac{d^2{\bf b}_{i}}{A_\perp} \,
 (-i)\int\frac{d^2 {\bf q}_{i}}{(2\pi)^2} \, {\cal A}_{el}(0,{\bf q}_{i}) \,
 e^{-i{\bf q}_{i}\cdot{\bf b}_{i}} \, 
(-i)\int\frac{d^2 {\bf q}_{i}^{\prime}}{(2\pi)^2} 
\, {\cal A}_{el}(0,{\bf q}_{i}^{\prime}) \, e^{-i{\bf q}_{i}^{\; \prime}
\cdot{\bf b}_{i}}  \; \cdots \rangle\; \nonumber \\
&=& \cdots (-1) \int \frac{d^2 {\bf q}_{i}}{(2\pi)^2} \,
\frac{d^2 {\bf q}_{i}^{\prime}}{(2\pi)^2} 
\frac{(2\pi)^2 \delta^2 ({\bf q}_{i} + 
{\bf q}_{i}^{\prime})}{A_\perp} 
{\cal A}_{el}(0,{\bf q}_{i}){\cal A}_{el}(0,{\bf q}_{i}^{\prime}) \; \cdots 
\;\; \nonumber \\
&=& 
\cdots (-1) \frac{\sigma_{el}}{A_\perp} 
\int \frac{d^2 {\bf q}_{i}}{(2\pi)^2} 
  \,\left[ \frac{1}{\sigma_{el}}
\frac{d\sigma_{el}}{d^2{\bf q}_i}  \right] 
 \; \int d^2 {\bf q}_{i}  \delta^2 ({\bf q}_{i} + 
{\bf q}_{i}^{\prime}) \; \cdots \, \;\; .
\label{travv}
\eea
The momentum transfers are constrained here to be equal and
opposite.
In Eq.~(\ref{travv}) we have used ${\cal A}_{el}(0,{\bf q}_{i})
= {\cal A}_{el}^\star(0,{\bf q}_{i})$ and 
${\cal A}_{el}(0,{\bf q}_{i}) = {\cal A}_{el}(0,-{\bf q}_{i})$.
\end{widetext}

\section{Amplitude iteration technique to second order in opacity
          for on-shell partons}
\label{bg2ord}

We illustrate the iteration of gluon emission amplitudes
to second order in opacity and calculate the Bertsch-Gunion 
case for  direct comparison to the general gluon emission result 
for soft 
scatterings~Eq.~(\ref{BGfull}). One should first recall that 
for the Bertsch-Gunion case 
\beq
G_0  =  {\bf 0} \;, 
\eeq{zero}
since the asymptotic on-shell quarks do not radiate 
without the medium-induced acceleration. For the two 
rank 1 classes, we apply once the direct and virtual 
insertion operators, $\htD_1$ from  Eq.~(\ref{didamit}) and  
$\htV_1$ from  Eq.~(\ref{vidamit}), to the 
incoming parton, Eq.~(\ref{zero}), and obtain:
\beqar
D_1 G_0 &=& -{\bf B}_1\, e^{i\omega_0 z_1} \, [c,a_1]  \;\;, \label{11} \\[1ex] 
V_1 G_0 &=& - \frac{C_A}{2}\,{\bf B}_1\,e^{i\omega_0 z_1}\, c  \label{22}  \;\;.
\eeqar{11apl} 
We note that the results are simple since only the new color current
${\bf B}_1$ contributes in the absence of a non-zero initial gluon amplitude.
%We have denoted in Eq.~(\ref{11apl})  $z_{ij} \equiv z_i-z_j,\; 
%\Delta z_i \equiv z_i-z_{i-1}$.

To second order in opacity, we build upon the amplitudes
from  Eqs.~(\ref{11}) and~(\ref{22}).  Some of the rank 2 classes  are 
obtained from rank 1 through relabeling, i.e. $ D_2 G_0 
\equiv D_1 G_0 (1\rightarrow 2),\; 
V_2 G_0 \equiv V_1 G_0 (1\rightarrow 2)$.
The rest are readily derived from  Eqs.~(\ref{11}) and~(\ref{22}) through
our iteration scheme Eqs.~(\ref{didamit}) and (\ref{vidamit}):  
\beqar 
D_2 D_1 G_0 &=& -{\bf B}_1\, e^{i\omega_0 z_1} \, a_2[c,a_1]  
-{\bf B}_2\, e^{i\omega_0 z_2} \, [c,a_2] a_1  
\nonumber \\[1.5ex] 
&& -{\bf B}_{2(12)}\, e^{i(\omega_0 z_2 -\omega_2 z_{21} )} 
\, [[c,a_2], a_1]  \;\;, 
\label{3} \\[1ex] 
V_2 D_1 G_0 &=& 
+ \frac{C_R+C_A}{2}\,  {\bf B}_1\, e^{i\omega_0 z_1} \, [c,a_1]  
\nonumber \\[1.5ex] 
&& -  \frac{C_A}{2}\,  {\bf B}_2\, e^{i\omega_0 z_2} \, c a_1  
\nonumber \\[1.5ex] 
&&+{\bf B}_{2(12)}\, e^{i(\omega_0 z_2 -\omega_2 z_{21} )} 
\,a_2 [[c,a_2], a_1]  \;\;, 
\label{4} \\[1ex] 
D_2 V_1 G_0 &=& 
- \frac{C_A}{2}\,  {\bf B}_1\, e^{i\omega_0 z_1} \, a_2 c  
+ \frac{C_R}{2}\,  {\bf B}_2\, e^{i\omega_0 z_2} \, [c, a_2]  
\nonumber \\[1.5ex] 
&&
-\frac{C_A}{2}\, {\bf B}_{2(12)}\, e^{i(\omega_0 z_2 -\omega_2 z_{21} )} 
\, [c,a_2]  \;\;, 
\label{5} \\[1ex] 
V_2 V_1 G_0 &=& 
+\frac{C_A(C_R+C_A)}{4}\, {\bf B}_1\, e^{i\omega_0 z_1} \, c  
\nonumber \\[1.5ex] 
&& + \frac{C_RC_A}{4}\,  {\bf B}_2\, e^{i\omega_0 z_2} \, c  
\nonumber \\[1.5ex] 
&&
-\frac{C^2_A}{4}\, {\bf B}_{2(12)}\, e^{i(\omega_0 z_2 -\omega_2 z_{21} )} 
\, c  \;\;. 
\label{66} 
\eeqar{2apl}  
With this explicit construction of the relevant classes of diagrams, we 
can compute the differential 
gluon emission  up to second order in the opacity expansion.

A necessary side step in the brute force approach 
is the evaluation of the color factors using the techniques 
in Refs.~\cite{Cvitanovic:1976am,Gyulassy:1999zd}.
We denote by $C_R$ the quadratic Casimir of the representation 
of the incident parton. For SU($N_c$), following the standard 
normalization for the generators  we have
\begin{equation}
C_F={N_c^2-1 \over 2N_c}, \qquad C_A=N_c \; .
\end{equation}
We recall that  in our notation  $T_c \equiv c$ for brevity. 
In the absence of  interactions we have only one color matrix 
associated with  the gluon  emission vertex from a hard scatter,
\beq
{\cal C}^{(0)}_1 = c  \; ,
\eeq{vaccol}
with 
\beq
\tr \; {\cal C}^{(0) \dagger }_1 {\cal C}^{(0)}_1  = C_R D_R \; .
\eeq{vacccolsq}
To first order in opacity we have  to consider the following
color factors
\begin{eqnarray}
& {\cal C}^{(1)}_1=  c  a_1 \; ,  \quad   
{\cal C}^{(1)}_2 =  \left  [ c , a_1 \right ] \; ,  \quad   
{\cal C}^{(1)}_3 = a_1 c \;, 
\label{1col}
\end{eqnarray} 
with
\beqar
&& \tr \; {\cal C}^{(1) \dagger }_1 {\cal C}^{(1)}_1  = C_R^2 D_R \; ,
\quad {\cal C}^{(1) \dagger }_2 {\cal C}^{(1)}_2  = C_R C_A D_R \; ,
\nonumber \\[1ex] 
&&  \tr \; {\cal C}^{(1) \dagger }_3 {\cal C}^{(1)}_3  = C_R^2 D_R \; ,
\nonumber \\[1ex] 
&& \tr \; {\cal C}^{(1) \dagger }_1 {\cal C}^{(1)}_2 = 
\frac{C_A}{2} C_R D_R  \; , 
\quad \nonumber \\[1ex] 
&&\tr \; {\cal C}^{(1) \dagger }_1 {\cal C}^{(1)}_3  = 
\left(C_R-\frac{1}{2}C_A\right) C_R D_R\;,
 \nonumber \\[1ex] 
&&\tr \; {\cal C}^{(1) \dagger }_2 {\cal C}^{(1)}_3  = 
\frac{C_A}{2}  C_R D_R\;.
\eeqar{1colsq} 
To second order in opacity we have:
\beqar
&& {\cal C}^{(2)}_1 = a_2 a_1 c\; , \quad 
{\cal C}^{(2)}_2 = a_2 
\left [ c , a_1 \right ] \; ,  \nonumber \\[1ex]
&&  {\cal C}^{(2)}_3  = \left[ c , a_2 \right] a_1 \; ,  \quad 
{\cal C}^{(2)}_4 =  c \, a_2 a_1  \; , \quad 
 \nonumber \\[1ex]
&&{\cal C}^{(2)}_5 = \left[  \left[ c , a_2 \right] , a_1 
\right] \; .
\eeqar{2orcol} 
The traces over the color factors yield:
\begin{eqnarray}
&& \tr\, {\cal C}_1^{(2)\dagger}{\cal C}^{(2)}_1 = 
 \tr\, {\cal C}_4^{(2)\dagger}{\cal C}^{(2)}_4 = 
C^3_R D_R\;,  \quad 
\nonumber \\[1ex]   
&& \tr\,{\cal C}^{(2)\dagger}_2{\cal C}^{(2)}_{2} 
 = \tr\,{\cal C}^{(2)\dagger}_3{\cal C}^{(2)}_{3} 
 =  C_A C^2_R D_R\;,  \quad
\nonumber \\[1ex]   
&& \tr\,{\cal C}^{(2)\dagger}_{5}{\cal C}^{(2)}_{5}=
      C^2_A C_R D_R\; ,  \nonumber \\[1ex]
&&  \tr\,{\cal C}^{(2)\dagger}_1{\cal C}^{(2)}_2 = 
- \tr\,{\cal C}^{(2)\dagger}_3{\cal C}^{(2)}_4 =
-\frac{1}{2}C_A C^2_RD_R\;, 
\quad  \nonumber \\[1ex] 
&& \tr\,{\cal C}^{(2)\dagger}_1{\cal C}^{(2)}_3 
= - \tr\,{\cal C}^{(2)\dagger}_2 {\cal C}^{(2)}_4 
\nonumber \\ 
&&  \hspace*{1.9cm}
= -\frac{1}{2} \left(C_R-\frac{1}{2}C_A\right)C_A C_R D_R\;,
\nonumber \\
&&  \tr\,{\cal C}^{(2)\dagger}_1{\cal C}^{(2)}_4
= \left(C_R-\frac{1}{2}C_A\right)^2  C_R D_R\;,
\nonumber \\
&& \tr\,{\cal C}^{(2)\dagger}_1{\cal C}^{(2)}_5  = 0\;,  
\nonumber \\ 
&& \tr\,{\cal C}^{(2)\dagger}_2{\cal C}^{(2)}_3 = 
\tr\,{\cal C}^{(2)\dagger}_2{\cal C}^{(2)}_5=
- \tr\,{\cal C}^{(2)\dagger}_4{\cal C}^{(2)}_5 
 \nonumber \\[1ex] 
&& \hspace*{1.9cm}  = -\frac{1}{4}C^2_A C_R D_R\;,
\quad  \nonumber \\[1ex] 
&& \tr\,{\cal C}^{(2)\dagger}_3{\cal C}^{(2)}_5=\frac{1}{2}C^2_A C_R\;, 
\quad 
\label{second}
\end{eqnarray} 
Although some of the color factors, such as ${\cal C}^{(1)}_3$ and 
${\cal C}^{(2)}_4$, do not appear in the amplitudes listed above, 
these will prove useful in the calculation of the realistic case 
of incoming partons that undergo hard scattering to produce 
final-state jets.

To first order in opacity,  the  radiation 
from either {\em quark} or {\em gluon} jets reads:
\beqar
k^+ \frac{N^{(1)}_g(BG)}{d k^+ d^2 {\bf k}} 
& \propto &   \frac{1}{D_R} \, 
\left \langle \; {\rm Tr}  \left[ (D_1G_0)^\dagger D_1 G_0 
 \right. \right. \nonumber \\[1ex]
&\ &  \hspace*{0.8cm}  
\left. \left. + (  (G_0)^\dagger  V_1 G_0  + h.c. ) \right] 
\; \right \rangle \;. \qquad
\eeqar{dn1amps}
In Eq.~(\ref{dn1amps}),  $\left \langle   \cdots  \right \rangle$ 
denotes the average over the momentum transfers from the 
medium. Including the phase space factor, the strong coupling 
constant $\alpha_s$ and  integrating over the longitudinal 
position of the scattering centers we find:
\beqar
k^+ \frac{N^{(1)}_g(BG)}{d k^+ d^2 {\bf k}}  & = &   
\frac{ C_R \alpha_s }{ \pi^2 }
 \int_0^L \frac{d \Delta z_1}{\lambda_g (z_1) } 
\;  \int d^2 {\bf q}_1 \; \nonumber \\[1ex] 
&& \hspace*{1cm} \times  \frac{1}{ \sigma_{el}(z_1) }
\frac{d \sigma_{el} }{d^2 {\bf q}_1}   
 \left[ \; | {\bf B}_1 |^2 \; \right] \; . \;\; \qquad
\eeqar{dn1ampsful}
Here,  we recover the Bertsch-Gunion result to one scattering 
center~\cite{Gunion:1981qs} and show that it is equivalent to 
first order in opacity. One important difference  between the
initial- and/or final-state radiation and  on-shell 
jets is that in the latter cases the first order in opacity 
does not lead to any interference/coherence effects.

The first non-trivial correction to the Bertsch-Gunion result 
can also be directly calculated. Using the amplitudes, 
Eqs.~(\ref{11}) - (\ref{2apl}), and the calculated color 
factors,   Eqs.~(\ref{1col}) - (\ref{second}), we find:  
\beqar
k^+ \frac{N^{(2)}_g(BG)}{d k^+ d^2 {\bf k}} 
& \propto &   \frac{1}{D_R} \,  \left \langle \;  {\rm Tr} 
\left[ \,  (D_2 D_1 G_0)^\dagger D_2 D_1 G_0  \right. \right.
\nonumber \\[1ex]
&&+  \; ( (D_2 G_0)^\dagger  D_2 V_1 G_0 + {\rm h.c.})  
\nonumber \\[1.ex]
&&+ \; (    ( D_1 G_0)^\dagger  V_2 D_1 G_0  + {\rm h.c.})  
\nonumber \\[1.ex]
&&  +  \;  ( (G_0)^\dagger  V_2 V_1 G_0 + {\rm h.c.} ) 
\nonumber \\[1.ex]
&& \left. \left. 
+  \;  (  (V_2 G_0)^\dagger  V_1 G_0  + {\rm h.c.} )  \,  \right] 
\;  \right \rangle  \; .
\eeqar{secordsch} 
The final result for the first non-trivial coherence 
correction reads:
\beqar
k^+ \frac{N^{(2)}_g(BG)}{d k^+ d^2 {\bf k}} 
& = &  \frac{ C_R \alpha_s }{ \pi^2 }
 \int_0^L \frac{d \Delta z_1}{\lambda_g (z_1) } 
\int_0^{L-\Delta z_1} \frac{d \Delta z_2}{\lambda_g (z_2) } 
\nonumber \\[1ex]
&& \hspace*{-1.5cm} \times  \int {d^2 {\bf q}_1}
  \frac{1}{ \sigma_{el}(z_1) }
\frac{d \sigma_{el} }{d^2 {\bf q}_1}   \;  
\int d^2 {\bf q}_2 \;   \frac{1}{ \sigma_{el}(z_2) }
\frac{d \sigma_{el} }{d^2 {\bf q}_2}   \nonumber \\[1ex]
 && \hspace*{-1.5cm} \ \times  \left[ \;  | {\bf B}_{2(12)} |^2  
-  | {\bf B}_1 |^2    +  2 {\bf B}_2 \cdot  {\bf B}_{2(12)}  
\cos (\omega_2 z_{21})  \;  \right]        \; . \nonumber \\[1ex]  
&&
\eeqar{secord}
In a medium with no sharp boundaries we can extend the 
integration limits as follows, $z(0) \rightarrow - \infty$, 
$z(L) \rightarrow + \infty$. The disappearance of the  
interactions is guaranteed by $\lambda_g(z_i) \rightarrow \infty$ 
as $\rho(z_i) \rightarrow 0$.  Direct comparison to the full 
recurrence solution, Eq.~(\ref{BGfull}), can now be made with 
consistent results, as expected.

\section{Amplitude iteration technique to second order in opacity for 
         initial-state radiation}
\label{is2ord}

The case of initial-state energy loss, as emphasized in the general 
derivation section, differs from the full Bertsch-Gunion solution 
by the presence of the additional hard scatter at position $L$. 
While the $t=-\infty$ state for the incoming 
on-shell parton remains the same, the hard acceleration yields an 
additional term:
\beq
{H} = \left( -\frac{1}{2} \right)^{ {N}_v(\vAi)}  
\, {\bf H} e^{i \omega_0 z_L} 
\; c T_{el}(\vAi) \; .  \quad
\eeq{zero1} 
In the absence of soft  interactions,
\beq
G_0 + H  = {\bf H} e^{i \omega_0 z_L} \; c \;\; .
\eeq{zero11}

The dependence of $H$ on the sequence of interactions preceding the 
hard scatter is implicit on our notation.
If we want to explicitly verify the general result to second order 
in opacity, we need to calculate only a few additional terms relative to
Appendix~\ref{bg2ord}. 
The relevant amplitudes are modified as follows, 
\beqar
D_1 G_0 + H &=& -{\bf B}_1\, e^{i\omega_0 z_1} \, [c,a_1]  
+ {\bf H} e^{i \omega_0 z_L} \, c a_1 
\;\;, \label{1}\qquad  \\[1ex] 
V_1 G_0 +H &=& - \frac{C_A}{2}\,{\bf B}_1\,e^{i\omega_0 z_1}\, c  
- \frac{C_R}{2}{\bf H} e^{i\omega_0 z_L} \, c 
\label{2}  \;\;, \qquad
\eeqar{1apl} 
and
\beqar 
D_2 D_1 G_0 +H &=& -{\bf B}_1\, e^{i\omega_0 z_1} \, a_2[c,a_1]  
\nonumber \\[1.5ex] 
&&-{\bf B}_2\, e^{i\omega_0 z_2} \, [c,a_2] a_1  
\nonumber \\[1.5ex] 
&& -{\bf B}_{2(12)}\, e^{i(\omega_0 z_2 -\omega_2 z_{21} )} 
\, [[c,a_2], a_1]  \nonumber \\[1.5ex] 
&& + {\bf H} e^{i \omega_0 z_L} \, c a_2 a_1  \;\;, 
\label{33} \\[1ex] 
V_2 D_1 G_0 + H &=& 
+ \frac{C_R+C_A}{2}\,  {\bf B}_1\, e^{i\omega_0 z_1} \, [c,a_1]  
\nonumber \\[1.5ex] 
&& -  \frac{C_A}{2}\,  {\bf B}_2\, e^{i\omega_0 z_2} \, c a_1  
\nonumber \\[1.5ex] 
&&+{\bf B}_{2(12)}\, e^{i(\omega_0 z_2 -\omega_2 z_{21} )} 
\,a_2 [[c,a_2], a_1] 
\nonumber \\[1.5ex] 
&& - \frac{C_R}{2}  {\bf H} e^{i \omega_0 z_L} \, c  a_1  \;\;, 
\label{44} \\[1ex] 
D_2 V_1 G_0 + H &=& 
- \frac{C_A}{2}\,  {\bf B}_1\, e^{i\omega_0 z_1} \, a_2 c  
\nonumber \\[1.5ex] 
&&+ \frac{C_R}{2}\,  {\bf B}_2\, e^{i\omega_0 z_2} \, [c, a_2]  
\nonumber \\[1.5ex] 
&&
-\frac{C_A}{2}\, {\bf B}_{2(12)}\, e^{i(\omega_0 z_2 -\omega_2 z_{21} )} 
\, [c,a_2]  
\nonumber \\[1.5ex] 
&& - \frac{C_R}{2}  {\bf H} e^{i \omega_0 z_L} \, c  a_2  \;\;, 
\label{55} \\[1ex]  
V_2 V_1 G_0 +H &=& 
+\frac{C_A(C_R+C_A)}{4}\, {\bf B}_1\, e^{i\omega_0 z_1} \, c  
\nonumber \\[1.5ex] 
&& + \frac{C_RC_A}{4}\,  {\bf B}_2\, e^{i\omega_0 z_2} \, c  
\nonumber \\[1.5ex] 
&&
-\frac{C^2_A}{4}\, {\bf B}_{2(12)}\, e^{i(\omega_0 z_2 -\omega_2 z_{21} )} 
\, c \nonumber \\[1.5ex] 
&& + \frac{C_R^2}{4}  {\bf H} e^{i \omega_0 z_L} \, c  
 \;\;, 
\label{6} 
\eeqar{22apl}
in comparison to  Eqs.~(\ref{11}) - (\ref{2apl}). 
Having also calculated the color factors for this case in 
Appendix~\ref{bg2ord}, we obtain
\begin{widetext}
\beqar
k^+ \frac{N^{(2)}_g(IS)}{d k^+ d^2 {\bf k}} 
& \propto &   \frac{1}{D_R} \,  \left \langle \;  {\rm Tr} 
\left[ \,  (D_2 D_1 G_0+H)^\dagger (D_2 D_1 G_0 +H) 
 +   ( (D_2 G_0+H)^\dagger  (D_2 V_1 G_0 +H ) + {\rm h.c.})  \right. \right.
\nonumber \\[1.ex]
&&+ \; (    ( D_1 G_0 + H)^\dagger  (V_2 D_1 G_0 +H) + {\rm h.c.})  
  +   ( (G_0+H)^\dagger  (V_2 V_1 G_0 + H) + {\rm h.c.} ) 
\nonumber \\[1.ex]
&& \left. \left. 
+    (  (V_2 G_0 + H )^\dagger  ( V_1 G_0 +H ) + {\rm h.c.} )  
\,  \right]  \;  \right \rangle  \; .
\eeqar{secordis} 
\end{widetext}
Evaluating the additional terms, arising from the interference of the 
hard gluon bremsstrahlung with the multiple Bertsch-Gunion sources, and 
showing explicitly that the term $\propto {\bf H}^2$ cancels, we obtain:
\begin{widetext}
\beqar
k^+ \frac{N^{(1)}_g(IS)}{d k^+ d^2 {\bf k}}  
& = &  \frac{ C_R \alpha_s }{ \pi^2 }
 \int_0^L \frac{d \Delta z_1}{\lambda_g (z_1) } 
%\nonumber \\[1ex]
%&& \hspace*{-1.5cm} \times  
\int {d^2 {\bf q}_1}  \frac{1}{ \sigma_{el}(z_1) }
\frac{d \sigma_{el} }{d^2 {\bf q}_1}   \;  
  \left[ \;  | {\bf B}_1 |^2   \; 
- 2  {\bf H} \cdot  {\bf B}_{1} \cos (\omega_0 z_{L1} )   \right]      
  \; .  \qquad
\eeqar{firordh} 
\beqar
k^+ \frac{N^{(2)}_g(IS)}{d k^+ d^2 {\bf k}}  
& = &  \frac{ C_R \alpha_s }{ \pi^2 }
 \int_0^L \frac{d \Delta z_1}{\lambda_g (z_1) } 
\int_0^{L-\Delta z_1} \frac{d \Delta z_2}{\lambda_g (z_2) }
%\nonumber \\[1ex]
%&& \hspace*{-1.5cm} \times  
\int {d^2 {\bf q}_1}  \frac{1}{ \sigma_{el}(z_1) }
\frac{d \sigma_{el} }{d^2 {\bf q}_1}   \;  
\int d^2 {\bf q}_2 \;   \frac{1}{ \sigma_{el}(z_2) }
\frac{d \sigma_{el} }{d^2 {\bf q}_2} \;    
\left[ \;  | {\bf B}_{2(12)} |^2  \right. 
\nonumber \\[1ex]
&& 
\;  \left. \;  -  | {\bf B}_1 |^2  +  2 {\bf B}_2 \cdot {\bf B}_{2(12)}  
\cos (\omega_2 z_{21})  - 2 {\bf H} \cdot  {\bf B}_{2(12)}  
\cos (\omega_0 z_{L2} + \omega_2 z_{21}) 
+ 2  {\bf H} \cdot  {\bf B}_{1} \cos (\omega_0 z_{L1} )   \;  \right]      
  \; .  \qquad
\eeqar{secordh} 
\end{widetext}
These results coincide with the general opacity expansion  series
for initial-state medium-induced non-Abelian 
bremsstrahlung, see Eq.~(\ref{ISfull}).

\end{appendix}

\end{document}